\documentclass[a4paper,fleqn]{cas-sc}
\usepackage[square,sort,comma,numbers,compress]{natbib}

\usepackage{bm}
\usepackage{amsmath}
\usepackage{blindtext}
\usepackage{float}
\usepackage{xfrac}
\usepackage{xcolor}
\usepackage{amssymb, graphicx, mathtools, url, hyperref,cancel}
\usepackage{subcaption}
\usepackage[nodisplayskipstretch]{setspace}
\usepackage{array}
\usepackage{multicol}
\usepackage{etoolbox}

\usepackage{nomencl}

\def\tsc#1{\csdef{#1}{\textsc{\lowercase{#1}}\xspace}}
\tsc{WGM}
\tsc{QE}
\tsc{EP}
\tsc{PMS}
\tsc{BEC}
\tsc{DE}

\begin{document}
\let\WriteBookmarks\relax
\def\floatpagepagefraction{1}
\def\textpagefraction{.001}

\shorttitle{Cislunar Review}    

\title[mode=title]{A comprehensive review on Cislunar expansion and space domain awareness}

\shortauthors{Brian Baker-McEvilly, Surabhi Bhadauria, et al.}  


\author[inst1]{Brian Baker-McEvilly}[orcid=0000-0001-8952-7236]
\ead{bakermcb@my.erau.edu}

\author[inst2]{Surabhi Bhadauria}[orcid=0000-0001-6890-4426]
\ead{sbhadaur@purdue.edu}

\author[inst1]{David Canales}[orcid=0000-0003-0166-2391]
\ead{canaled4@erau.edu}

\author[inst2]{Carolin Frueh}[orcid=0000-0002-0240-5509]
\ead{cfrueh@purdue.edu}

\affiliation[inst1]{organization={Embry-Riddle Aeronautical University, Aerospace Engineering Department},
            addressline={}, 
            city={Daytona Beach},
            postcode={}, 
            state={Florida},
            country={USA}}

\affiliation[inst2]{organization={Purdue University, School of Aeronautics and Astronautics},
            addressline={}, 
            city={West Lafayette},
            postcode={}, 
            state={Indiana},
            country={USA}}

\begin{abstract}
The Cislunar region is crucial for expanding human presence in space in the forthcoming decades. This paper presents a comprehensive review of recent and anticipated Earth-Moon missions, and ongoing space domain awareness initiatives. {\color{black}An introduction to the dynamics as well as periodic trajectories in the Cislunar realm is presented. Then, a review of modern Cislunar programs as well as smaller missions are compiled to provide insights into the key players pushing towards the Moon. Trends of Cislunar missions and practices are identified, including the identification of regions of interest, such as the South Pole and the Near-rectilinear halo orbit. Finally, a review of the current state and short-comings of space domain awareness (SDA) in the region is included, utilizing the regions of interest as focal points for required improvement. The SDA review is completed through the analysis of the Artemis 1 trajectory.} 
\end{abstract}


\makenomenclature
\renewcommand{\nomname}{Acronyms}
\setlength{\nomitemsep}{-\parsep}
\renewcommand\nomgroup[1]{%
  \item[\bfseries
  \ifstrequal{#1}{T}{Terminology}{%
  \ifstrequal{#1}{M}{Missions \& Agencies}{}}%
]}
\renewcommand{\nompreamble}{\begin{multicols}{2}}
\renewcommand{\nompostamble}{\end{multicols}}

\maketitle
\nomenclature[T]{\(CR3BP\)}{Circular Restricted Three-Body Problem}
\nomenclature[T]{\(SDA\)}{Space Domain Awareness}
\nomenclature[T]{\(SSA\)}{Space Situational Awareness}
\nomenclature[T]{\(IM\)}{Intuitive Machines}
\nomenclature[T]{\(CLPS\)}{Commercial Lunar Payload Services}
\nomenclature[T]{\(DSN\)}{Deep Space Network}
\nomenclature[T]{\(NRHO\)}{Near-Rectilinear Halo Orbit}
\nomenclature[T]{\(LLO\)}{Low Lunar Orbit}
\nomenclature[T]{\(DRO\)}{Distant Retrograde Orbit}
\nomenclature[T]{\(BCR4BP\)}{Bi-Circular Restricted Four Body Problem}
\nomenclature[T]{\(LTV\)}{Lunar Terrain Vehicles}
\nomenclature[T]{\(TO\)}{Task Orders}
\nomenclature[T]{\(CLEP\)}{Chinese Lunar Exploration Program}
\nomenclature[T]{\(GEO\)}{Geosynchronous Earth Orbit}
\nomenclature[T]{\(HEO\)}{High Earth Orbit}
\nomenclature[T]{\(EKF\)}{Extended Kalman Filter}
\nomenclature[T]{\(UKF\)}{Unscented Kalman Filter}

\nomenclature[T]{\(SCR\)}{Shackleton Connecting Ridge}
\nomenclature[T]{\(SNR\)}{Signal-to-Noise Ratio}
\nomenclature[T]{\(ELFO\)}{Elliptical Frozen Lunar Orbit}
\nomenclature[T]{\(ISRU\)}{In-situ Resource Utilization}

\nomenclature[M]{\(ISRO\)}{Indian Space Research Organization}
\nomenclature[M]{\(JAXA\)}{Japan Aerospace Exploration Agency}
\nomenclature[M]{\(NASA\)}{National Aeronautics and Space Administration}
\nomenclature[M]{\(SELENE\)}{Selenological and Engineering Explorer}
\nomenclature[M]{\(LRO\)}{Lunar Reconnaissance Orbiter}
\nomenclature[M]{\(LCROSS\)}{Lunar Crater Observation and Sensing Satellite}
\nomenclature[M]{\(GRAIL\)}{Gravity Recovery and Interior Laboratory}
\nomenclature[M]{\(LADEE\)}{Lunar Atmosphere and Dust Environment Explorer}
\nomenclature[M]{\(CAPSTONE\)}{Cislunar Autonomous Positioning System Technology Operations and Navigation Experiment}
\nomenclature[M]{\(KPLO\)}{Korea Pathfinder Lunar Orbiter}
\nomenclature[M]{\(SLIM\)}{Smart Lander to Investigate the Moon}
\nomenclature[M]{\(PRIME\)}{Polar Resource Ice Mining Experiment}
\nomenclature[M]{\(TRIDENT\)}{The Regolith and Ice Drill for Exploration of New Terrains}
\nomenclature[M]{\(LUPEX\)}{Lunar Polar Exploration}
\nomenclature[M]{\(VIPER\)}{Volatiles Investigating Polar Exploration Rover}

\printnomenclature[2.25cm]

\doublespacing
\section{Introduction} \label{sec:Intro}
\begin{figure}[b!]
    \hfill{}
    \centering
    \includegraphics[width=10cm]{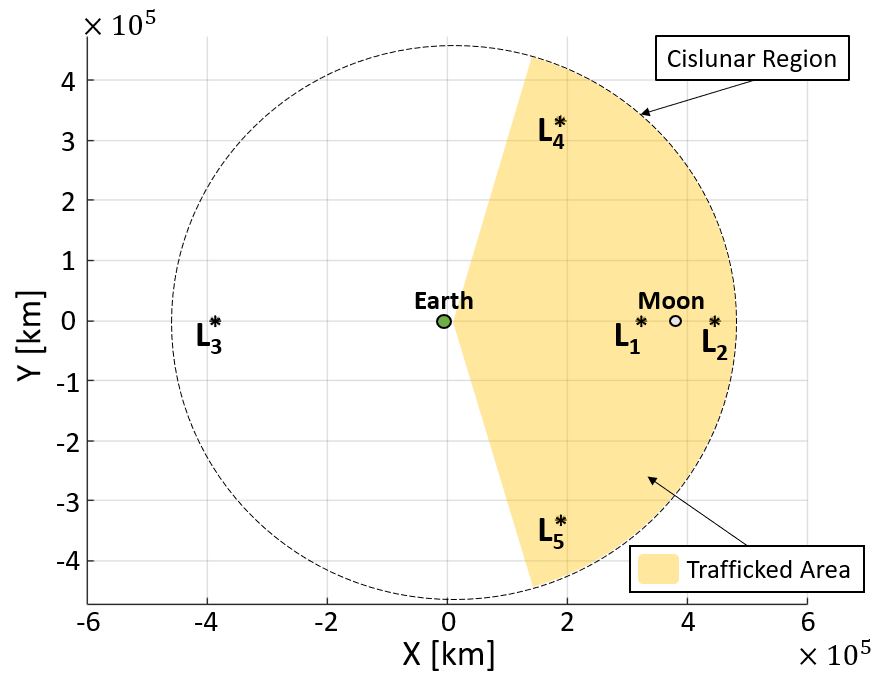}\hfill{}
    \caption{\label{fig:CisRegion}A representation of the Cislunar region, within the circle, and the space traficked area shaded in yellow.}
\end{figure}

In the next decade, the Cislunar region will host over 30 missions, many of which contain multiple payloads and experiments traveling to the Moon's surface and into Lunar orbit~\cite{Brian, Johnson2022, ArtOverview}. It can be recognized that the Cislunar realm holds intrinsic value for scientific, commercial, and military applications as numerous entities have begun to invest resources into the expansion and utilization of the Cislunar realm~\cite{RoadMap}. The Cislunar region is considered to be space in which the gravitational effect of the Sun, Earth, and Moon have significant influence over a spacecraft~\cite{Primer}. The region can be further refined to typically trafficked areas within the vicinity of the Moon. A visual depiction of the Cislunar region is displayed in Figure \ref{fig:CisRegion}. Cislunar missions aim to test Cislunar trajectories~\cite{CAPOverview}, map the Moon's surface~\cite{VIPER}, and test resource harvesting methods from the Lunar soil, also called regolith~\cite{PRIME}. The overarching goal for the majority of these Cislunar missions is to support the development of a Lunar base that can sustain long-term presence of humans. Furthermore, a successfully pioneered Lunar base would perpetuate Cislunar traffic, requiring efficient resupply and transport of both humans and resources. Given the revitalization of large scale Lunar programs, the goal of long-term human occupation is fueled by the desire to utilize Lunar resources in space missions, use the Moon as a foothold for missions extending into deeper space, and test technology. Overall, the Cislunar region is set to be the stage for an expansion of human influence into space on a scale never before seen. 


{\color{black}Knowledge of all operations within a region of space, or space domain awareness (SDA), is paramount to the continued success and utilization of the covered region. Missions situated about or near the Earth are accessible through Earth's extensive observational infrastructure, including ground radars and optical instruments. This infrastructure offers missions near the Earth the availability of space object monitoring, detecting, and tracking services across a high volume of objects. These SDA capabilities about the Earth are significant and enable missions to operate safely and efficiently throughout the shared space. However, the extensive observational infrastructure on Earth struggles to sufficiently cover all of Cislunar space to the same extent it covers around Earth due to the distance and challenging observational conditions \cite{fruehcislunar}. Some Earth-based facilities, such as NASA's deep space network (DSN) \cite{DSNref}, are capable of reaching into Cislunar space, but handle a significantly lower volume of objects compared to near-Earth networks. The shortcoming of adequate Cislunar SDA capabilities, in conjunction with the increasing volume of Cislunar traffic, stands as a hindrance to the utilization and expansion of Cislunar space. Therefore, a review of Cislunar SDA is warranted to encompass the current steps being investigated to bridge this cap in SDA, as well as outline challenges and shortcomings of the current infrastructure. Although Cislunar space is seeing a significant increase in traffic, it is a vast region of space that will not all be put to use. Providing sufficient SDA across the entirety of Cislunar space would primarily be a moot service as only key regions of interest will be occupied by spacecraft and need these services. Space domain awareness services in these key regions, in turn, enables more missions to occupy the space safely by offering key services and lowering the bar of entry to operate within the region. Due to this relationship, knowledge of Cislunar missions and Cislunar SDA are not inherently independent topics, and offer insight into one another. Knowledge of Cislunar missions further highlights trends and motivations for this new wave of Cislunar expansion, offering insight into the future of Cislunar space. Thus, the goal of this paper is to complete a review of the current state of missions within the Cislunar space and the current state of the region's space domain awareness capabilities to support these Cislunar missions. This review aims to address two themes within the context of Cislunar operations: (a) where are missions traveling and what are their motivations, and (b) what is the current state and limitations of space domain awareness in the region to support such missions?}

{\color{black}The work approaches the topics through an integrative review process, completing a synthesis of available information and then providing insight into the topics through demonstrating applications or drawing relevant conclusions. The technical sources utilized in this review are selected based upon them recently being published and acknowledged by the community, written by reputable authors, or published by authoritative organizations within the aerospace community. As a first source meeting this criteria, reputable journals and conferences are referenced. Alternatively, published dissertations of graduates from schools reputable in the field are sourced. As this manuscript is a review of current topics in the Cislunar region, newly written sources are a focus. However, some older sources used to build the foundation of a well researched topic are employed. A key topic of this work is the analysis of modern Cislunar missions, including recent and future missions. The review of recent modern missions includes those that experienced crashes or failures, as they still offer valuable insight into the modern landscape of Cislunar space. The review of future missions can be difficult, as the space industry is volatile and details of future mission operations may be altered. Thus, to ensure the accuracy and longevity of this manuscript, future missions elaborated upon must be funded and well into development, or backed by large scale, authoritative organization such that the likelihood of drastic changes or cancellations is minimal. Furthermore, sensitive future mission information, such as launch dates, are not included as they are likely to change. Typically, missions that have not flown will not have readily available technical conference or journal publications. In the event sources for Cislunar mission are not available within conferences or journals, NASA's Space Science and Data Coordinate Archive is referenced.}

The paper is formatted such that a thorough review is presented. {\color{black}A simple background in Earth-Moon system dynamics models is first explained in Section \ref{sec:Dynamics}. The section begins with the circular restricted three-body problem and elaborates on periodic trajectories in that dynamical framework.} To continue, an extensive summary of current and future Cislunar programs, missions, and key players are explored in Section \ref{sec:Missions}. The missions reviewed in Section \ref{sec:Missions} are used in Section \ref{sec:ROI} to provide insight on Cislunar mission trends and key regions of interest on the Lunar surface and in space. {\color{black}An understanding of common mission motivations and the establishment of regions of interest in the Cislunar realm are of paramount importance as it produces key locations to center efforts of space domain awareness about. The review is completed with an overview of the current methods, applications, and limitations of space domain awareness across the region in Section \ref{sec:CislunarSDA}. The SDA review is completed with an visibility analysis of the Artemis I trajectory and final conclusions on Cislunar SDA.} 

\section{Dynamical Background} \label{sec:Dynamics}
{\color{black}An introductory understanding of the Earth-Moon dynamical models and their implications offer an important background to the Cislunar region. In particular for this review, this is key for the missions and SDA concepts that are discussed in subsequent sections. First, a well-accepted method of modeling motion in the Earth-Moon system, the circular restricted three-body problem (CR3BP) is summarized \cite{Canales, Gupta, Grebow}. Then, periodic trajectories within the context of the Earth-Moon CR3BP are introduced. The concepts introduced shape the dynamics of activities within the Cislunar space, affecting where missions may go} {\color{black} \cite{GovCis}. This offers essential context for discussions about Cislunar missions and SDA, further detailed in Sections \ref{sec:Missions} and \ref{sec:CislunarSDA}.}

\begin{figure}[h!]
    \hfill{}
    \centering
    \includegraphics[width=12cm]{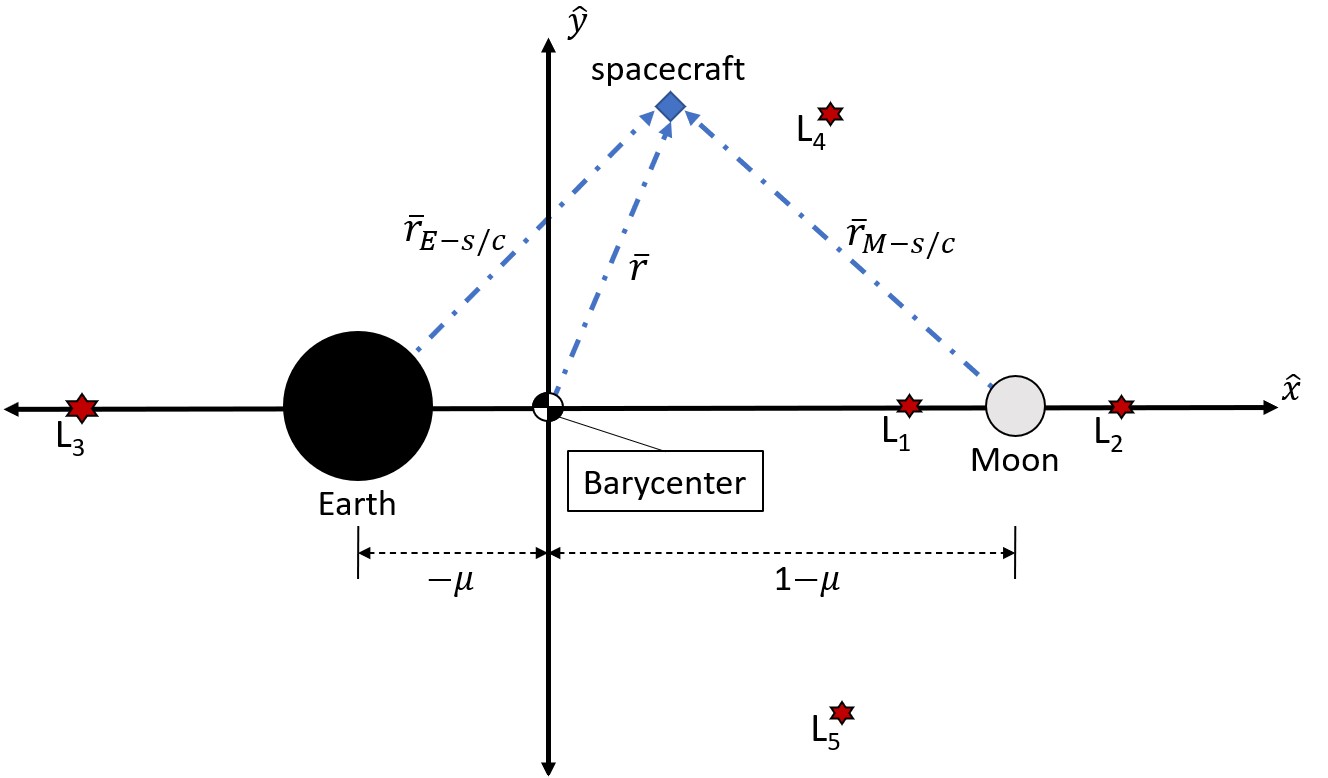}\hfill{}
    \caption{\label{fig:CR3BP}Schematic of the Earth-Moon CR3BP system.}
\end{figure}

In the Earth-Moon system, the gravitational influence of both the Earth and the Moon are significant to the motion of a spacecraft. The CR3BP describes motion under the influence of two primary gravitational bodies. In such a model, no analytical solution has yet been derived and numerical methods shall be utilized. The CR3BP considers three bodies: a large and small primary (Earth and Moon, respectively) as well as a spacecraft of negligible mass. The primaries follow circular orbits around their common barycenter, identified using the mass parameter of the system. {\color{black}The mass parameter ($\mu$) is defined by:}
\begin{align}
\label{eq:MassParameter}
\mu=\frac{m_{M}}{m_{E}+m_{M}},
\end{align}
where $m$ is the mass and the subscripts $E$ and $M$ refer to the Earth and Moon, respectively. It is convenient to present the equation of motion for the CR3BP system in the Earth-Moon rotating reference frame in which the origin is located at the barycenter of the system, the positive $\hat{x}-axis$ points towards the Moon, the positive $\hat{z}-axis$ points perpendicular to the Earth-Moon orbital plane, and finally the $\hat{y}-axis$ completes the right-handed system. Furthermore, the CR3BP system is nondimensionalized in order to reduce the scale of the Earth-Moon system, reducing strain on applied numerical methods. To nondimensionalize the CR3BP system, characteristic length, time, and mass are utilized: characteristic mass corresponds to the sum of the Earth's and Moon's mass; characteristic length is the average distance between the Earth and Moon; and the characteristic time is selected to ensure the mean motion of the Earth and Moon are equal to unity. The state of the spacecraft in the CR3BP is denoted by $\bar{r}=[x,y,z]^T$ and velocity by $\dot{\bar{r}}=[\dot{x},\dot{y},\dot{z}]^T$, where the superscript $T$ represents the transpose of a vector and a dot above a term denotes its derivative with respect to time. Thus, the motion of a spacecraft in the CR3BP is governed by the following non-dimensional equations:
\begin{align}
\label{eq:CR3BP}
\ddot{x}=2\dot{y}+\frac{\partial U^*}{\partial x},
\qquad
\ddot{y}=-2\dot{x}+\frac{\partial U^*}{\partial y},
\qquad
\ddot{z}=\frac{\partial U^*}{\partial z}
\end{align}
in which $U^*$ is the pseudo-potential function of the system: 
\begin{align}
\label{eq:Potential}
U^*=\frac{1-\mu}{\vert \vert \bar{r}_{E-s/c} \vert \vert}+\frac{\mu}{\vert \vert \bar{r}_{M-s/c} \vert \vert}+\frac{1}{2}(x^2+y^2)
\end{align}
where ${\bar{r}_{E-s/c}}$ is the position vector from the Earth to the spacecraft, ${\bar{r}_{M-s/c}}$ is the position vector from the Moon to the spacecraft, and the double bars on each side of a vector denotes magnitudes~\cite{Brian, Canales} (Figure \ref{fig:CR3BP}). In the CR3BP, there exists points where the gravitational influence of the primaries as well as rotational effects from the Earth-Moon rotating frame become null. These are referenced as Lagrange/libration points ($L_1 - L_5$) and their stability properties are often leveraged for trajectory design \cite{Grebow}.

\begin{figure}[h!]
\centering
\begin{subfigure}{.45\linewidth}
    \includegraphics[width=1\linewidth]{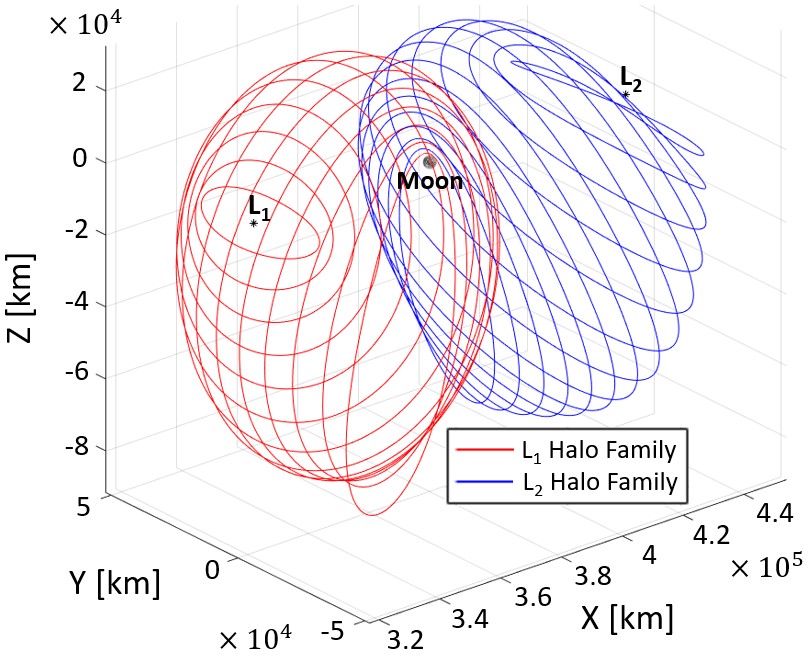}
    \centering
    \caption{$L_1$ \& $L_2$ Southern Halo Orbit Family.}\label{fig:Halo}
\end{subfigure}
    \hfill
\begin{subfigure}{.42\linewidth}
    \includegraphics[width=1\linewidth]{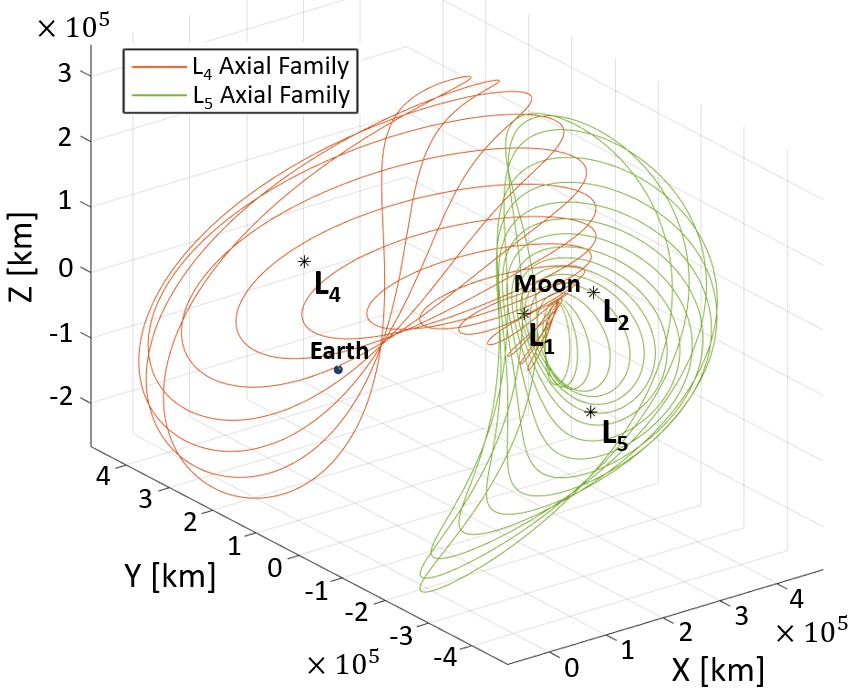}
    \centering
    \caption{$L_4$ \& $L_5$ Axial Orbit Family.} \label{fig:Axial}
\end{subfigure}
\bigskip
\begin{subfigure}{.45\linewidth}
    \includegraphics[width=1\linewidth]{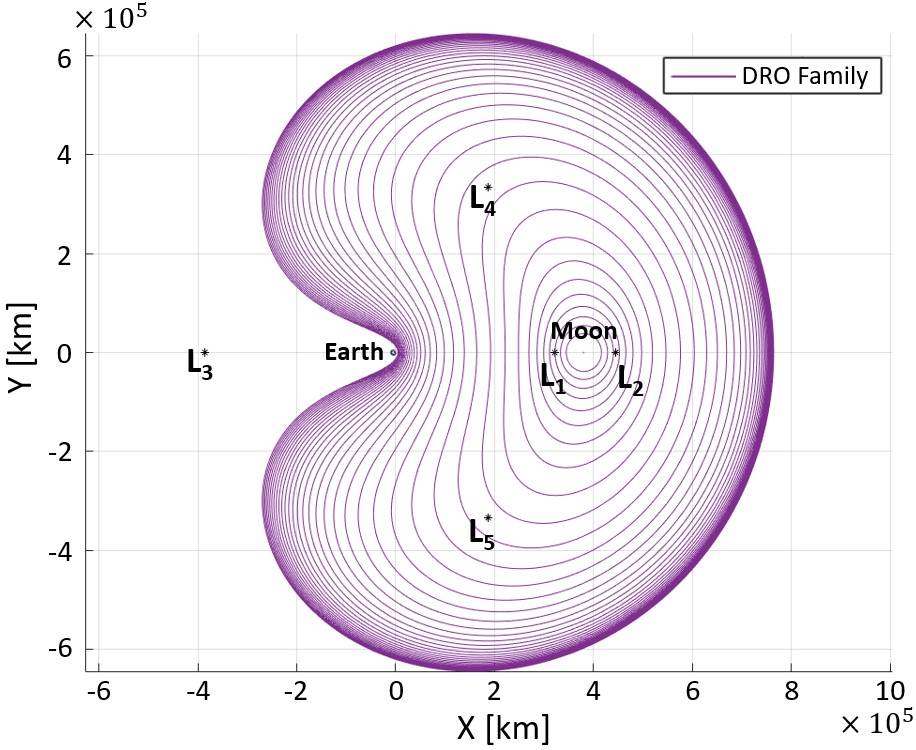}
  \centering
  \caption{Distant Retrograde Orbit Family.}\label{fig:DROFamily}
\end{subfigure} 
\hfill
\begin{subfigure}{.43\linewidth}
    \includegraphics[width=1\linewidth]{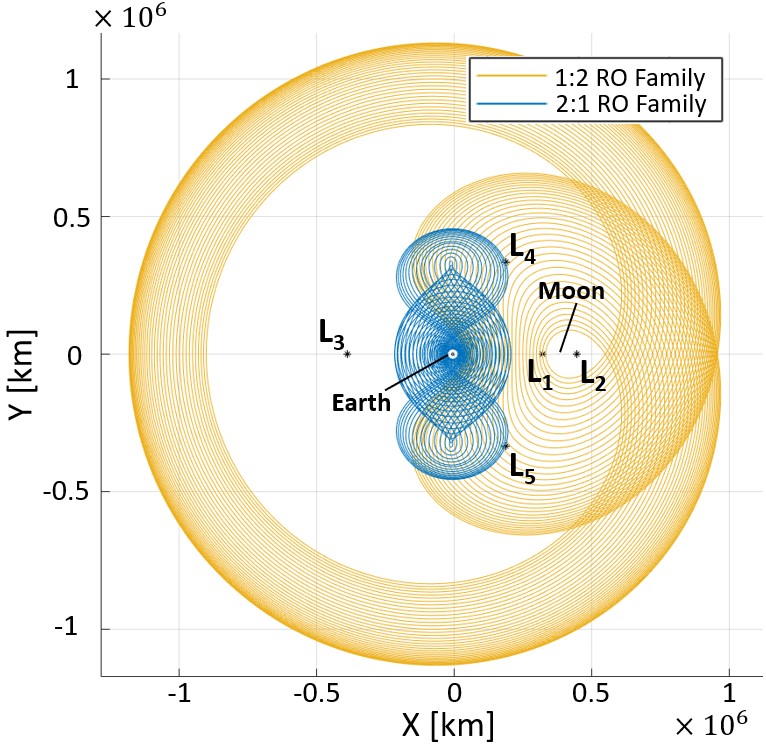}
    \centering
    \caption{$2:1$ \& $1:2$ Resonant Orbit Family.}\label{fig:ResFamily}
\end{subfigure}
\caption{Sample Orbit Families.}
\label{fig:OrbitFamilies}
\end{figure}

The stability properties of the libration points allows for periodic motion to occur in their vicinity. This periodic motion is categorized into distinct families of orbits according to stability and geometry~\cite{Grebow}. The families of orbits in the CR3BP consist of trajectories bifurcating from one another, connected through stability properties. These trajectories are obtained numerically, but some have seen real-world mission applications. Orbit families offer a myriad of possibilities to navigate the complex dynamics of the Earth-Moon system. Four orbit families are presented here to provide an overview of key trajectories for Cislunar exploration (Figure \ref{fig:OrbitFamilies}): the $L_1$ \& $L_2$ southern halo family, the $L_4$ \& $L_5$ axial family, the 2-D distant retrograde orbit (DRO) family, and a 2:1 \& 1:2 resonant orbit family. Beyond the selected options, there are a diverse multitude of families that are well-reviewed~\cite{Grebow,OrbitFamily}. Note that within an orbit family, an infinite number of periodic orbits exist. Therefore, the orbits selected in Figure \ref{fig:OrbitFamilies} are chosen to demonstrate the general evolution of the family, and do not represent the family in totality. 

A desirable property of the halo orbit family is that the orbits offer continuous line of sight of various latitudes of the Moon and the Earth. The Chinese Lunar mission Chang'E 4 deployed the Queqiao relay satellite into an $L_2$ halo orbit. The satellite acted as a relay from the Moon to Earth for Chang'E 4, as Chang'E 4 landed on the far-side of the Moon, having no direct communication back to Earth~\cite{Queqiao}. A subset of the halo orbit family called the near-rectilinear halo orbits (NRHO) lay extremely close to the Moon and are highly inclined. An NRHO is the proposed location of a long-term orbiting station called Gateway, mostly due to the low propulsion needed to remain in the orbit given its inherit stability~\cite{GateBook}. The $L_4$ \& $L_5$ axial orbit family demonstrates how unique periodic motion can be in the Cislunar region. Such a family possessing a wide range of motion allows access to a large portion of the Cislunar region. Additionally, $L_4$ \& $L_5$ orbit families have been investigated to utilize the Earth and Moon for occultations, allowing for high-quality observations of the Solar corona and other astrophysical processes~\cite{LunarOccultations}, as well as investigated to house observational spacecraft \cite{ObsSc}. The DRO family is a stable family that contains orbits close to $L_1$ \& $L_2$. The large and relatively simple motion of the family allows for efficient transportation throughout the Cislunar region~\cite{Art1TrajAnalysis}. Finally, resonant orbits are a result of the natural relationship between the periods of orbits. A resonant orbit is described by an integer ratio of periods between two bodies (resonance ratio). In context of the CR3BP, the ratio is represented by $p:q$, where $q$ is the number of orbits the Moon completes in the same time a spacecraft completes $p$ orbits. Note that both $p$ and $q$ are close integer approximations of the periods. The $2:1$ orbit family has been investigated to house observational spacecraft, as it is a stable trajectory that expands over large reaches of Cislunar space~\cite{GuptaSSA}. The unstable resonant orbits present in the CR3BP are often used to complete transfers between different areas of the Earth-Moon system (i.e., $2:3$ Resonance), while stable resonant orbits offer quasi-periodic trajectories that tour large portions of the Cislunar region (i.e., $3:1$ Resonance)~\cite{Sadaka2023}. Overall, the CR3BP and periodic trajectories in Cislunar space define the foundation in which missions may function in the region. {\color{black} Higher-fidelity models exist of the Earth-Moon system, such as models incorporating non-spherical gravity in the CR3BP \cite{J2Pert}, orbit and attitude coupling in the CR3BP model \cite{SE3}, the bi-circular restricted four body problem (BCR4BP) that considers the Sun's gravitational influence \cite{de2022comparisons, oshima2022multiple}, and the ephemeris model that considers planetary ephemerides (positions of celestial objects at corresponding epochs) for numerous bodies \cite{HowellNRHO}. However, the CR3BP and periodic trajectories in Cislunar space define the foundation in which missions may evolve in the region and provide the context needed for this review.}

\section{Cislunar Missions} \label{sec:Missions}
Space-faring countries have begun to revitalize large scale Lunar programs at a rate not seen since the Apollo era. {\color{black}Following the success of Chang'E 5, China has proposed three Lunar missions out of its Chinese Lunar Exploration Program~\cite{ChanOver}. In late 2023, Russia renewed its Lunar program Luna with the launched Luna 25~\cite{Luna25NA}, 37 years after the launch of Luna 24 by the Soviet Union.} Many other space faring countries are developing smaller scale Cislunar missions over the next decade~\cite{Johnson2022}. At the forefront of the Cislunar expansion, the United States created the Artemis program~\cite{ArtIG} as well as the Commercial Lunar Payload Services (CLPS) initiative~\cite{CLPSSum}. The larger scale Cislunar programs (Artemis, Chang'E, Luna) aim towards the establishment of long-term Lunar bases, while small programs aim to carry out research, experiments, and tests that will support the larger programs or promote commercial companies. {\color{black}In consideration of this increased Cislunar interest, the Cislunar region is a focal point for space faring organizations over the next decade.}

{\color{black} The Cislunar region possesses a rich history of missions that form the foundation of Cislunar knowledge. Table \ref{PastMissionsTable} catalogs key Lunar missions from the 21$^{st}$ century, showcasing the efforts of various organizations contributing diverse payloads, components, and mission services. Only the country or organization directly affiliated with each mission is listed. While these missions have provided invaluable scientific data and insights, the table highlights just one key achievement per mission to maintain conciseness.}
\begin{table}[h!]
\caption{{\color{black}Summary of key 21$^{st}$ century Cislunar missions.}}
\label{PastMissionsTable}
\begin{tabular}{lllll}
\hline
\textbf{Mission} & \textbf{Launch Date} & \textbf{Affiliation}    & \textbf{Execution} & \textbf{A Key Achievement}                                            \\ \hline
SELENE \cite{SELENE}          & Sep, 2007            & Japan                    & Success            & Analyzed material composition of the Moon                             \\
Chang'E 1 \cite{ChanOver}       & Oct, 2007            & China                    & Success            & China's first time reaching Lunar orbit                               \\
Chandrayaan 1 \cite{Chan1}   & Oct, 2008            & India                    & Success            & India's first mission to Lunar orbit                                  \\
LRO \cite{LRO}             & Jun, 2009            & USA                      & Success            & Created a 3D map of the Lunar surface                                 \\
LCROSS  \cite{LCROSSv2}         & Jun, 2009            & USA                      & Success            & Confirmed water in Lunar south poles                      \\
Chang'E 2 \cite{ChanOver}       & Oct, 2010            & China                    & Success            & Returned high-res images of Lunar surface                      \\
GRAIL \cite{GRAIL}           & Sep, 2011            & USA                      & Success            & Mapped the gravity about the Moon                                     \\
LADEE \cite{LADEE}           & Sep, 2013            & USA                      & Success            & Examined the Lunar atmosphere and surface       \\
Chang'E 3 \cite{ChanOver}       & Dec, 2013            & China                    & Success            & China's first soft landing on the Moon                                \\
Chang'E 5 \cite{ChanOver}       & Oct, 2014            & China                    & Success            & China's first Lunar sample return                                     \\
Queqiao \cite{Queqiao}          & May, 2018            & China                    & Success            & Relay satellite operating in Earth-Moon L2                            \\
Chang'E 4 \cite{ChanOver}       & Dec, 2018            & China                    & Success            & First soft landing on far side of the Moon                             \\
Beresheet \cite{Beresheet}       & Feb, 2019            & Israel                   & Failure               & Israel's first Lunar missions                                         \\
Chandrayaan 2    & Jul, 2019            & India                    & Failure               & India's first Lunar landing attempt                        \\
CAPSTONE \cite{CAPOverview}        & Jun, 2022            & USA                      & Success            & First spacecraft to operate in an NRHO                                \\
KPLO (Danuri) \cite{KPLODynamics}    & Aug, 2022            & South Korea              & Success            & South Korea's first Lunar mission                                     \\
Artemis 1 \cite{Art1Traj}       & Nov, 2022            & USA                      & Success            & First flight of Orion spacecraft                                      \\
Lunar Icecube \cite{IceCube}   & Nov, 2022            & USA                      & Success            & Investigated concentration of Lunar water \\
Hakuto-R \cite{HakNA}        & Dec, 2022            & ispace (Japan)            & Failure               & First landing attempt by ispace                           \\
Chandrayaan 3 \cite{ChanNA}   & Jul, 2023            & India                    & Success            & India's first soft landing on Lunar surface                           \\
Luna 25 \cite{Luna25NA}         & Aug, 2023            & Russia                   & Failure               & Russia's return to Moon since 1976                         \\
SLIM \cite{SLIM}            & Sep, 2023            & Japan                    & Success            & Demonstrated Japanese precision landings                             \\
TO2-AB \cite{TO2ABNA}          & Jan, 2024            & Astrobotics (USA)         & Failure               & Astrobotic's first Lunar mission                                      \\
TO2-IM \cite{TO2IMNA}          & Feb, 2024            & IM (USA) & Succsess       & First soft landing by a commercial lander                              \\
Queqiao 2 \cite{Queqiao2}       & Mar, 2024            & China                    & Success            & Relay satellite operating in an ELFO                                  \\
Chang'E 6 \cite{ChangE6}        & May, 2024            & China                    & Success            & Lunar South pole sample return                                        \\ \hline
\end{tabular}
\end{table}

\subsection{Artemis \& Gateway} \label{sec:Artemis}
\begin{figure}[t!]
    \hfill{}
    \centering
    \includegraphics[width=11cm]{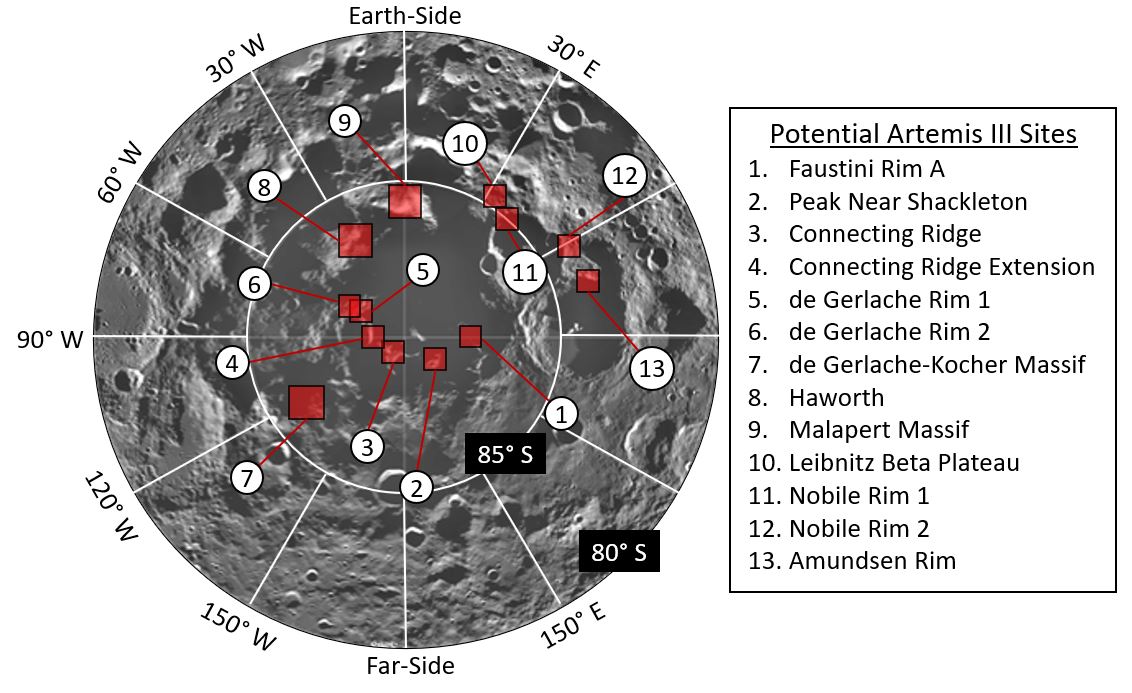}\hfill{}
    \caption{\label{fig:Art3Sites}Candidate landing sites for Artemis III.}
\end{figure}

The push into Cislunar space by NASA is seen through its new Lunar exploration program, Artemis, and a new Lunar orbital station, Gateway. The Artemis program is tasked with the goal of expanding human influence into the Cislunar region, as well as establishing a long-term base on the Moon~\cite{ArtOverview}. Artemis is a five-staged program; it begins with the testing of next generation rockets and ends with the establishment of a Lunar base. Artemis I successfully launched in November of 2022, with the primary goal of testing the Orion rocket and the habitat module that will carry humans in future Artemis launches~\cite{ArtRef}. The module flew in a partial DRO around the Moon and returned back to Earth safely 25 days later~\cite{Art1Traj, Art1TrajAnalysis}. The next Artemis mission, Artemis II, will execute a similar mission with the addition of a crew and new experiments, completing a Lunar fly-by and return to Earth~\cite{Art2}. {\color{black} Artemis II is well into the implementation stages of development, with notable milestones in April of 2023 with the selection of the Artemis crew and the beginning of ground systems testing in 2024.} Once the Orion rocket and human habitat module prove capabilities for Lunar transport via Artemis I and II, the next sequence of missions sets the stage for a Lunar base. Artemis III will transport humans into Cislunar space and down to the Lunar South Pole. This will be the first return of humans on the Lunar surface since Apollo 17 in 1972~\cite{ArtIG,ArtOverview}. Artemis III is designed to be the first mission to dock onto the Gateway station, utilizing the onboard modules to traverse to/from the Lunar surface. In the event Gateway is not completed by the time Artemis III is ready to launch, Artemis III will be designed so that it transports to and from the Lunar surface independently of Gateway~\cite{Art3Over}. As of late 2022, the potential landing sites for Artemis III have been announced, all falling in the South Pole region~\cite{Art3Sites}. Figure \ref{fig:Art3Sites} displays the candidate landing sites and corresponding geographic name.
{\color{black} Further progress of Artemis III  includes the evolution of the Lunar terrain vehicle (LTV) that Artemis III astronauts will use to traverse the Lunar surface. In April of 2024, NASA awarded Intuitive Machines {\color{black}(IM)}, Lunar Outpost, and Venturi Astrolab feasibility task orders in which they will propose concepts of the LTV that meets NASA's requirements in competition for the final order of LTVs.} The fourth leg of the Artemis program will dock onto Gateway upon its arrival in the Cislunar region. Artemis IV astronauts will deploy to the surface of the Lunar South Pole from Gateway and begin establishing a Lunar base~\cite{ArtSites}. Finally, Artemis V will complete the first set of missions for the program, whose mission objective will adjust as the program evolves. However, given the current trend of Artemis III and Artemis IV, it will likely visit the South Pole region of the Moon to further progress Artemis's goal of establishing a Lunar base in the South Pole~\cite{ArtSPFocus}. {\color{black} As the later Artemis missions follow the extensive development of the earlier stages of the Artemis program, they will advance efficiently with NASA's full attention as their launch dates approach.} Over the course of five missions, the Artemis program will establish a long-term base in the Lunar South Pole. 

\begin{figure}[t!]
    \hfill{}
    \centering
    \includegraphics[width=10cm]{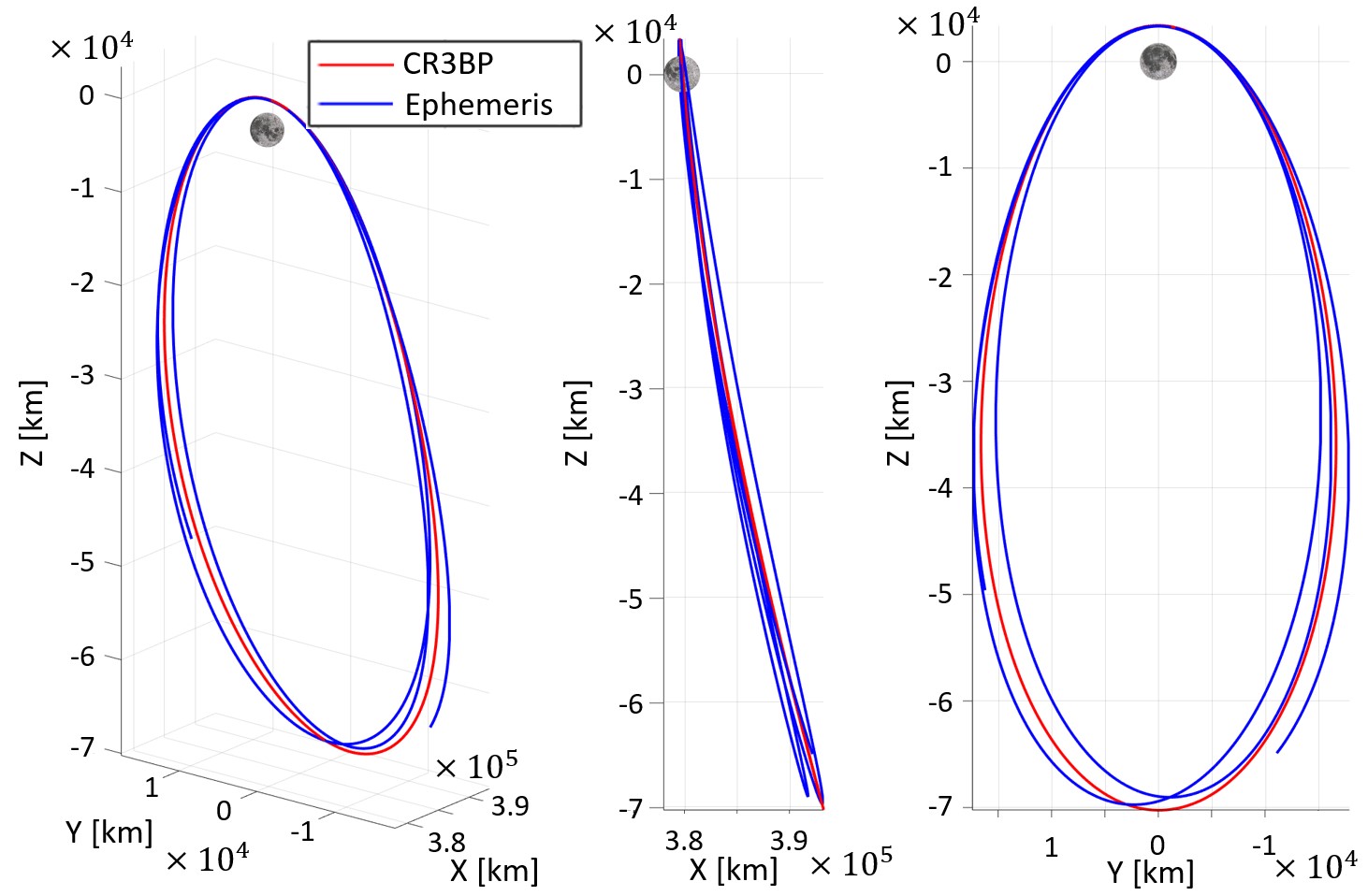}\hfill{}
    \caption{\label{fig:GatewayNRHO}Gateway's NRHO in a higher-fidelity ephemeris model defined on epoch July $5^{th}$, $2023$ presented in barycentric Earth-Moon reference frame. In ephemeris, Moon is the central body, while Earth and Sun are perturbing bodies.}
\end{figure}

The United States, Canada, Japan, and many European countries are contributing to the development of different pieces of the Gateway orbital station~\cite{GatewayInernational}. The Gateway station is a multi-national piece of Cislunar infrastructure that will enable the continuous exploration and settlement of the Cislunar region. Gateway will be a pivotal piece of infrastructure to the Artemis program, acting as a staging point for the Artemis missions, specifically Artemis IV and V, and many other missions wishing to utilize Gateway. Astronauts will deploy back and forth between the Lunar surface and Gateway using available onboard systems~\cite{GateBook}. {\color{black} The progress of Gateway is well underway, the power and propulsion element (PPE) module as well as the Lunar I-hab module completed there preliminary design reviews in late 2021, while the Habitation and Logistic Outpost (HALO) module of the station completed its critical design review in late 2022.} The modules onboard Gateway will continue to grow in complexity and versatility as more are shipped to the Moon throughout its lifetime, becoming an aggregation point for space technology. This culmination of space infrastructure and technology will be utilized by NASA to further reach towards Mars~\cite{MarsForward}. Gateway is strategically placed to act as the Cislunar center for communication relay between the Lunar surface and Earth. At the same time, it will provide observations of the Moon and deeper portions of Cislunar space, as well as provide additional mission support to the Lunar South Pole~\cite{GatewayInernational}. Gateway will follow a southern NRHO with a 9:2 resonance ratio, an orbit tested by the Cislunar Autonomous Positioning System Technology Operation and Navigation Experiment (CAPSTONE)~\cite{CAPOverview}. A large portion of Gateways capabilities are enabled by the dynamics of the 9:2 NRHO (Figure \ref{fig:GatewayNRHO}). 
The 9:2 NRHO has demonstrated favorable stability properties in the CR3BP and retains those properties when transitioned to higher-fidelity models~\cite{HowellNRHO}. The inherit stability is seen in the estimated $10 \: \sfrac{m}{s}$ velocity adjustment required every year to remain in orbit \cite{GatewayNRHO}, a low requirement for the region. The 9:2 NRHO further enables Gateway's success due to (a) both the moderate thrust needed to enter the orbit from Earth as well as to leave the orbit to travel to the Lunar surface, (b) the constant line of sight to the Earth, and (c) the minimal eclipse time~\cite{GatewayNRHO}. The latter produces stable thermal environments, preventing drastic changes in temperatures due to traveling in and out of sunlight for extended periods of time. The prolonged exposure to the Sun also prevents the need for large power systems. Gateway and the Artemis program are clear examples of large scale funding, planning, and infrastructure that the United States is placing in the Cislunar region.

\subsection{Commercial Lunar Payload Services} \label{sec:CLPS}
\begin{figure}[t!]
    \hfill{}
    \centering
    \includegraphics[width=11cm]{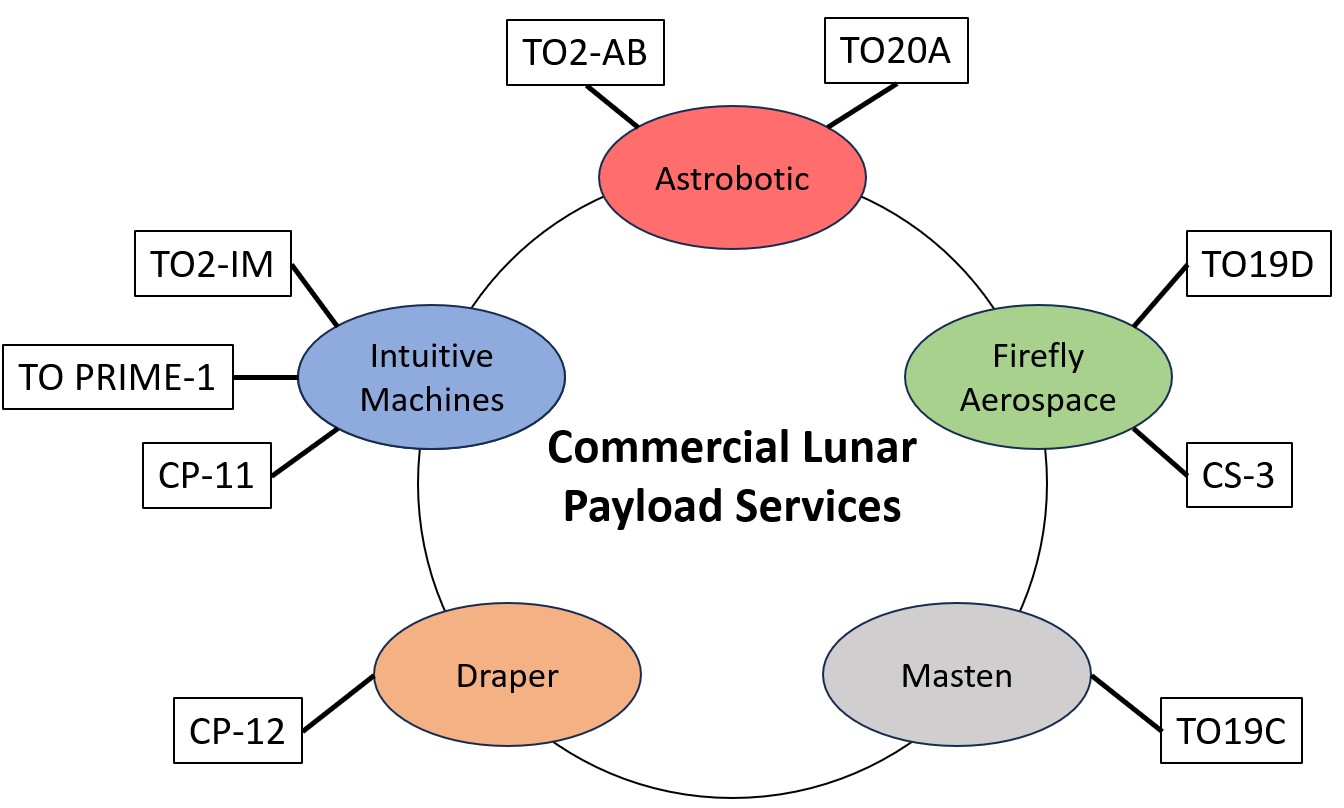}\hfill{}
    \caption{\label{fig:CLPSSummary}{\color{black}Summary of awarded CLPS missions.}}
\end{figure}

The Commercial Lunar Payload Services (CLPS) is a NASA initiative created to promote the commercial industry in Cislunar expansion, while continuing to place NASA experiments in space (Figure \ref{fig:CLPSSummary}). The CLPS program outlines mission requirements through task orders (TO), identifying payloads and payload delivery locations. Commercial companies bid and submit proposals to develop a lander that completes a task order. The winning proposal is then given the mission task order and funding is awarded to continue development. By the end of 2023, eleven CLPS missions are planned and nine have been awarded to companies~\cite{CLPSSum}. Intuitive Machines has been awarded three task orders (TO2-IM, TO PRIME-1, TO CP-11): with the goal to deliver payloads to different areas of the Cislunar region using its Nova-C lander~\cite{TO2IMNA, PRIME1NA, CP11NA}. Astrobotics has won two task orders (TO2-AB, TO20A): deploying the Peregrine lander to deliver TO2-AB and Griffin lander to deliver TO20A~\cite{VIPER}. Masten Space Systems will utilize the Xelene lander to complete the order TO19C~\cite{TO19CNA}. The CLPS orders, TO19D and TO CS-3 are to be carried out by Firefly Aerospace via the Blue Ghost lander~\cite{TO19DNA, TO19DPayload, CS3NA}. Draper rounds out the awarded task orders, delivering payloads under the CP-12 task order \cite{CP12NA}. The remaining two CLPS missions, CP-21 and CP-22, have yet to be awarded to a commercial company. {\color{black} Each of the awarded proposal for a CLPS task order undergo a rigorous review process in competition for the task order. Thus, upon acceptance of the task order and initial funding, the projects are well into development. Furthermore, for the companies that have flown CLPS missions, their landers are already flight tested and are primarily complete.} {\color{black}The first CLPS launch of Astrobotics T02-AB Peregrine lander suffered a fuel leak and was unable to complete its trip to the Moon, deorbiting into Earth's atmosphere after 10 days. Following Astrobotics failure, Intuitive Machine's IM-1 mission achieved a soft landing onto the Moon but fell onto its side after a landing strut collapsed. However, IM-1 was still operational and carried out briefly operated payloads that were unaffected by the tip over of the lander.}

Each CLPS lander will carry NASA and commercial payloads into predetermined locations in Cislunar space and on the Lunar surface. Many of the NASA payloads onboard the CLPS landers are experiments and tests that will provide data to inform decisions in the Artemis program. Across all CLPS missions, at least one experiment is present to study Lunar regolith or Lunar weather/surface conditions. Additionally, CLPS missions TO2-IM, T0 PRIME-1, TO20A, TO19C, will all carry payloads designed to investigate the presence of water on the Lunar surface~\cite{PRIME, TO2IMNA, TO19CNA, TO20ANA}. Most notably, the Polar Resource Ice Mining Experiment (PRIME) on-board PRIME-1, as well as Volatiles Investigating Polar Exploration Rover (VIPER) on-board TO20A, are two high-profile experiments that will investigate water on the Lunar surface. {\color{black} PRIME-1 is anticipated to land on Shackleton connecting ridge (SCR), while VIPER will explore the region around Mons Mouton.} Both experiments utilize the Regolith and Ice Drill for Exploration of New Terrains (TRIDENT) to penetrate one meter into the Lunar and extract samples. The PRIME payload will be mounted on the Nova-C lander used for the PRIME-1 CLPS mission. The TRIDENT drill will extend from the lander and begin to drill, searching for water~\cite{PRIME}. VIPER is a rover designed to traverse the Lunar surface for 100 days, sampling different portions of the Lunar soil for water~\cite{VIPER}. Although the majority of payloads aboard the CLPS landers aim to either study Lunar regolith, surface conditions, or water deposits, other experiments will also be carried out. For example, experiments aboard the TO19C mission and TO2-IM will study surface plume interactions with Lunar landers, and experiments aboard the TO19D and TO2-AB  will test Cislunar communication systems. As a result, the CLPS program contributes a significant chunk of Cislunar traffic over the next decade and paves the way for the commercial Cislunar industry.

\subsection{Chang'E \& Luna} \label{sec:Change}
The United States is not the only nation revitalizing large scale Lunar programs. Russia's Roscosmos space agency has announced the {\color{black}next missions of the Luna program}. Additionally, China has planned additional Cislunar missions to its Chinese Lunar Exploration Program (CLEP), also called the Chang'E program~\cite{ChanProgram}. Luna and Chang'E follow a similar approach to Cislunar expansion as Artemis, with the end goal to establish a Lunar base in the South Pole region of the Moon~\cite{ChanOver}. {\color{black}The modern Russia Lunar program was begun with the launch of Luna 25 in 2023}. Its goal was to bring an assortment of payloads to the South Pole, {\color{black}including a Lunar robotic arm that will be used to collect regolith samples to investigate regolith composition~\cite{Luna25Locations}. However, Luna 25 crashed on the surface of the Moon about a week after launch. Following} Luna 25, Luna 26 is planned to be an orbiter around the Moon while Luna 27 and Luna 28 will land in the South Pole region of the Lunar surface~\cite{Luna27Landing, Johnson2022}. 

China plans on continuing its Cislunar program after the success of Chang'E 5 Lunar sample return~\cite{Cha5Recon} through three additional missions for the Chang'E program. The Chang'E 6 Lunar lander will touch down in the Aitken Basin, located in the South Pole on the far-side of the Moon~\cite{ChanLocations}. Chang'E 6 will be the nation's second Lunar sample return while studying the South Pole region. Data collected from Chang'E 6 samples will provide China information that will influence the future South Pole locations of Chang'E 7 and Chang'E 8. The Chang'E 7 and Chang'E 8 missions will continue to contribute to the establishment of a Chinese Lunar base in the South Pole~\cite{ChanProgram}. The Chinese Lunar base will differ from the United States in that China plans to implement a large role of autonomous robots in the station, allowing the station to not always require astronauts on the Moon to function~\cite{ChanOver}. {\color{black} It is important to note that the development of Luna and Chang'E is difficult to track and monitor to the same detail as Artemis, due to the policies and information sharing practices of respective countries, and these programs must be monitored closely for accurate developments.} The United State's Artemis program is the largest scale ongoing Lunar program, but other influential countries will have a significant role in Cislunar space. The relaunch of the Luna program is still in infancy, but the Chang'E program is keeping pace and will have vigorous development in the near future.

\subsection{Small Scale Missions} \label{sec:Other}
Outside of the larger scale Lunar programs, many other countries and organizations are sending missions into the Cislunar region. These agencies are placing spacecraft into the region via ride shares off of larger missions or through the design of an independent mission. Many of these missions are the beginning of Cislunar programs or are companies aiming to expand commercial influence in the Cislunar region. The Lunar IceCube mission, Lunar Trailblazer, and Lunar Pathfinder missions ride share to the Cislunar region to be subsequently deployed into orbit. Morehead State University led the 6U Lunar IceCube Cubesat project that launched onboard the Artemis I mission. The purpose of Lunar IceCube is to investigate the composition and age of Lunar regolith. It will supply the future Artemis missions valuable data for selecting mission locations. Lunar IceCube flies a low-Lunar orbit (LLO), with a $100$ km periapsis and $5000$ km  apoapsis, at a high inclination to service the South Pole region~\cite{IceCube}. Another college-lead Cislunar Cubesat project is the Lunar Trailblazer designed by CalTech. {\color{black}The mission will sit as an additional payload on-board the TO PRIME-1 CLPS mission.} The goal of Trailblazer is to investigate the distribution and formation of Lunar water. Trailblazer will fly in a LLO, at around $100$ km in altitude~\cite{TrailOver}. {\color{black} The Trailblazer payload is nearly complete with the integration of its final key sensor, the Lunar thermal mapper, in August of 2023.} Lastly, the Lunar Pathfinder mission is slated to be a payload on-board the CS-3 CLPS mission. Pathfinder aims to test and establish Cislunar communications and navigational services. It will offer services, such as data relay, to missions attending the region. Pathfinder will orbit in an elliptical frozen Lunar orbit, a unique form of LLO that retains its periodic dynamics for extended periods of time~\cite{Pathfinder}. {\color{black} The Lunar Pathfinder missions is in advanced stages of planning, completing its critical design review in early 2023.}

In the next decade, the Cislunar region will see historical achievements for new agencies sending missions into the area. Following the crash of the Chandrayaan-2 mission in 2019, the Indian Space Research Organization (ISRO) launched Chandrayaan-3 as a follow-up in 2023. Chandrayaan-3 (Chan-3), {\color{black}landed on the Moon nearby the Chandrayaan-2 crash site, marking India's first successful missions to the surface of Moon~\cite{ChanNA}}. Next, the Hakuto-R 1 is a mission carried out by "ispace", a commercial space agency based in Japan. The Hakuto-R lander crashed into the Atlas crater in April 2023, losing contact with ispace. The lander carried an assortment of payloads, including the Rashid Rover and the Japanese Lunar Excursion Vehicle Rover~\cite{HakNA}. The Hakuto-R lander notes the first attempt of a commercial space agency to land on the Moon, representing a company instead of a nation. Exploration of Cislunar space is also fostering new international missions as well, with the first joint space mission between ISRO and Japan Aerospace Exploration Agency (JAXA). The Lunar Polar Exploration mission (LUPEX) is a joint mission between India and Japan to place a spacecraft in the Lunar South Pole to study the presence of water~\cite{LUPEXLand, LUPEXOver}. Finally, Korea Pathfinder Lunar Orbiter (KPLO) marks South Korea's first mission into the Cislunar region. The orbiter launched in August of 2022, carrying 5 payloads, and inserted itself into its science orbit in December of 2022. Such a mission will remain in a LLO as it studies the Cislunar environment and the Lunar surface~\cite{KPLODynamics}. The Chandrayaan-3, Hakuto-R, LUPEX, and KPLO missions aim to make history in the Cislunar region for their respective countries/agencies. The entrance of new entities into the Cislunar region further demonstrates the value the region holds for future space exploration. {\color{black} Table \ref{FutureMissionsTable} summarizes the future high-profile missions attending the Cislunar region. Please note that this table includes the missions discussed that are in the advanced stages of planning. Similar to Table \ref{PastMissionsTable}, many countries and organizations possess pivotal roles in these missions, but only the country/organization that is directly affiliated to the mission is listed.}
\begin{table}[h!]
\caption{{\color{black}Summary of key future Cislunar missions.}}
\label{FutureMissionsTable}\begin{tabular}{llll}
\hline
\textbf{Mission}  & \textbf{Affiliation} & \textbf{Location} & \textbf{A Key Objective}                                                                 \\ \hline
Gateway \cite{GatewayInernational}          & USA                  & NRHO              & Establish orbital checkpoint for surface operation                                       \\
Artemis II \cite{Art2}       & USA                  & Fly-by            & Confirm operation of all Orion systems with a crew                                       \\
Artemis III \cite{Art3Over}      & USA                  & South Pole        & Develop Artemis Lunar base                                                               \\
Artemis IV \cite{ArtIG}       & USA                  & South Pole        & Develop Artemis Lunar base                                                               \\
Artemis V \cite{ArtIG}        & USA                  & South Pole        & Develop Artemis Lunar base                                                               \\
Chang'E 7 \cite{ChanOver}        & China                & South Pole        & Develop Chinese Lunar base                                                               \\
Chang'E 8 \cite{ChanOver}        & China                & South Pole        & Develop Chinese Lunar base                                                               \\
Luna 26 \cite{Johnson2022}          & Russia               & South Pole        & Establish Russian presence on Lunar surface                                              \\
Luna 27 \cite{Luna27Landing}          & Russia               & South Pole        & Establish Russian presence on Lunar surface                                              \\
TO PRIME-1 \cite{PRIME1NA}       & IM (USA)             & SCR               & Analyze regolith composition and water contents             \\

TO20A \cite{TO20ANA}            & Astrobotics (USA)    & Nobile Region     & Explore south pole and sample regolith for water                                         \\

CP-11 \cite{CP11NA}            & IM (USA)             & Reiner Gamma      & Investigate surface environment about Lunar swirls                                       \\
CP-12 \cite{CP12NA}            & Draper (USA)         & Schrodinger Basin & Study Lunar seismic activity and heat flow                                \\
TO19C \cite{TO19CNA}            & Masten (USA)         & Haworth Crater    & Research regolith mineralogy and volatiles \\

TO19D \cite{TO19DNA}            & Firefly (USA)        & Mare Crisium      & Examine the Moon's magnetic environment                           \\

CS-3 \cite{CS3NA}             & Firefly (USA)        & Lunar far-side    & Study deep space using low/radio frequencies           \\

Lunar Pathfinder  \cite{Pathfinder} & EU                   & LLO               & Establish network of data relays about the Moon                                          \\
Lunar Trailblazer \cite{TrailOver} & USA                  & LLO               & Map distribution of water in Lunar surface regolith                                      \\
LUPEX \cite{LUPEXOver}            & Japan \& India       & South Pole        & Detailed exploration of South pole                                                       \\ \hline
\end{tabular}
\end{table}

\subsection{{\color{black}Cislunar Mission Trends and Regions of Interest}} \label{sec:ROI}
{\color{black}The rapid expansion of Cislunar space is inevitable as the large scale Lunar programs continue to unfold and the commercial industry is incentivized to efficiently carry Lunar payloads into the region. At a minimum, NASA will always provide a demand for Cislunar transportation that commercial companies must meet, with the CLPS initiative acting as the vessel of this relationship. As companies such as Intuitive Machines, Astrobotics, Firefly Aerospace, and others achieve a profitable model of a Cislunar delivery service, they will continue to optimize and perfect transportation methods. This system of advancement will lower the bar of entry into the region, allowing more missions to attend and occupy the space.} 

{\color{black}The commercial industry's expansion will lower the bar of entry for other missions and also promote the healthy operation of the large scale programs such as Artemis. This is demonstrated by the many CLPS payloads that are being used to collect pivotal information in support of Artemis. A multitude of experiments will be delivered to Cislunar space, but two interests persist through numerous payloads: landing technology and plumes, as well as Lunar water studies and utilization. The ability for astronauts to harvest resources on the Moon will relieve the required water needed to be launched with missions and provide some sustainability to a Lunar base. Water is paramount to astronauts survival and is possibly a source of fuel and oxidizer when decomposed. The CLPS mission TO PRIME-1 and the Viper rover onboard TO 20A will both carry drills in order to collect samples and study the presence of water within the South Pole \cite{PRIME}. These missions also attempt to demonstrate in-situ resource utilization (ISRU), or the capability of water to be extracted from the regolith. In addition, Astrobotics TO 2-AB mission and TO 19C carry an infrared spectrometer to identify water volatiles under the surface.}

{\color{black}The number of active missions in the region is significantly rising, but many locations in Cislunar space and on the Lunar surface will remain uninhabited. The large scale Lunar programs, such as Artemis and Chang'E, are progressive programs in which earlier missions pave the way for later missions. This progression further concentrates missions into specific regions, as seen by the numerous CLPS missions. Through understanding the motivation and location of previous and upcoming Cislunar missions, key regions of interest in Cislunar space emerge. These regions provide valuable insight into the precise locations of Cislunar growth. Furthermore, these locations will require the focus of SDA applications, as opposed to barren regions. Two analysis are completed: an analysis of regions of interest on the Lunar surface and in Cislunar space.} 

\subsubsection{The Lunar Surface}
\begin{figure}[h!]
    \hfill{}
    \centering
    \includegraphics[width=13cm]{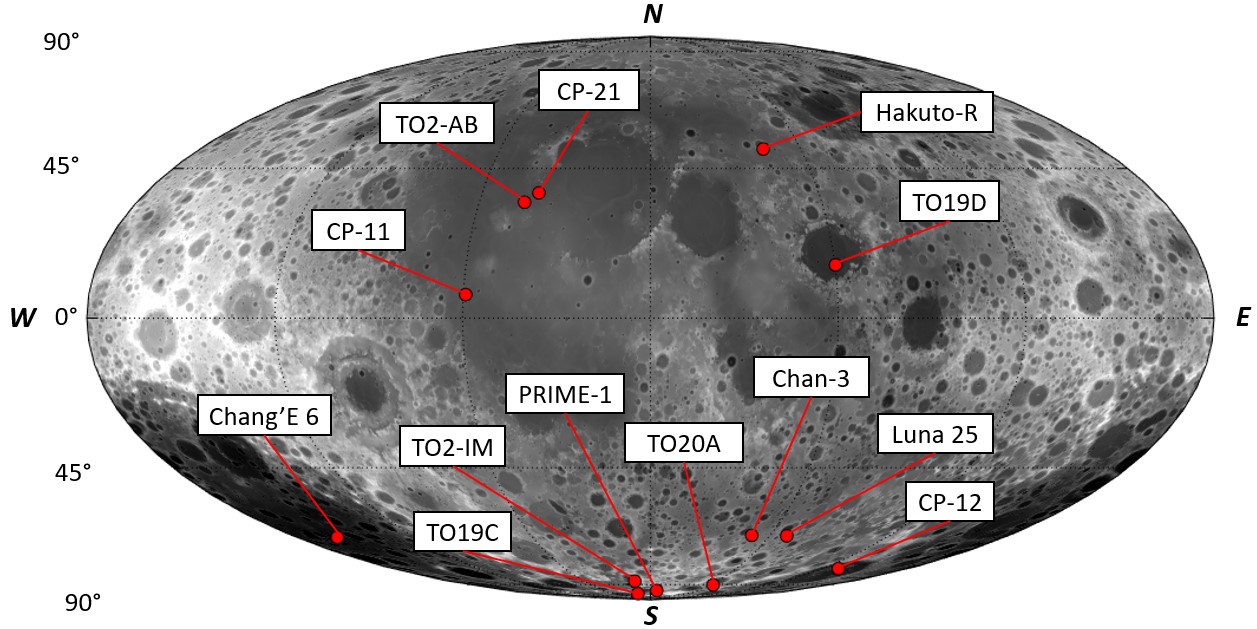}\hfill{}
    \caption{\label{fig:LunarMap}Lunar surface map of missions.}
\end{figure}

The Lunar surface is first investigated for key regions of interest. The locations of previous and future Lunar missions (including failed missions) are plotted onto the Lunar surface to provide insight into the populated areas of the Lunar surface (Figure \ref{fig:LunarMap}). The Lunar surface mapping does not depict every mission landing on the Lunar surface in upcoming years, as many missions either keep landing information confidential or rely on earlier mission data to determine landing sites. However, two key regions clearly emerge from Figure \ref{fig:LunarMap}: Earth-side craters and the South Pole. 

Earth-side craters will house six upcoming Cislunar missions. The Earth-side of the Moon offers a relatively safe experimental area compared to other regions like the South Pole or far-side as the Earth-side always is in direct line of sight of the Earth. Direct line-of-sight to Earth eases communication and surveillance challenges of missions as they may be seen from Earth continuously given proper station placement on Earth. {\color{black}Typically, mission operations on the Lunar surface will be carried out during the long Lunar days in which power generation is enabled via solar panels. In conjunction with direct line-of-sight to Earth, the long Lunar days provide sufficient illumination of mission operations in Earth-side craters for a significant duration of time.} Earth-side craters are a region of interest on the Lunar surface, allowing for a relatively safe environment on the Lunar surface to conduct experimentation and to test new technologies, as seen by the many missions attending the region.

\begin{figure}[b!]
    \hfill{}
    \centering
    \includegraphics[width=10cm]{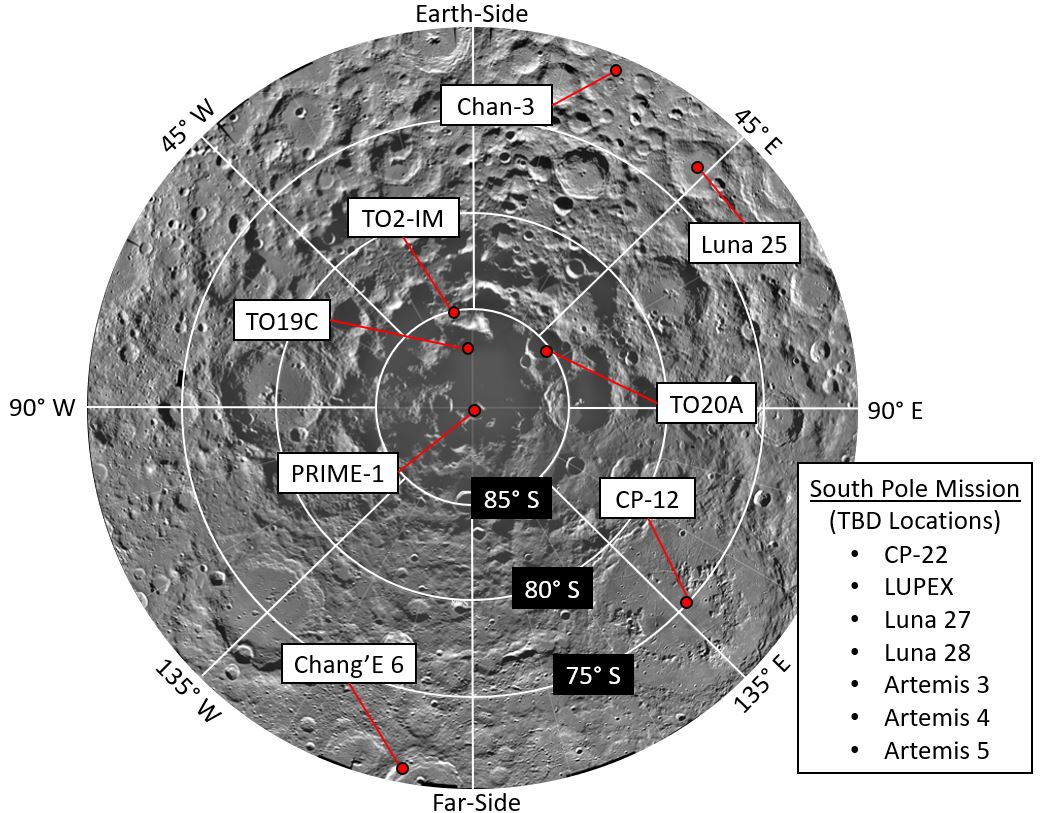}\hfill{}
    \caption{\label{fig:SouthPole}Identified landing sites in South Pole region of interest mission map.}
\end{figure}

The Lunar South Pole is the intended site of the Artemis Lunar base and a significant portion of Cislunar traffic. A map focused on the missions within the South Pole region of interest is presented in Figure \ref{fig:SouthPole}. 
An important geographical property of the region is that there are permanently shadowed craters in the area. Water exists in these South Pole craters, as the sunlight does not reach the surface to evaporate the water collected in the regolith. In 2009, the Lunar Crater Observation and Sensing Satellite (LCROSS) was purposefully crashed into the Lunar South Pole, confirming presence of water in the plume of Lunar regolith~\cite{LCROSS}. Water is a pivotal resource for Cislunar expansion, {\color{black}as it is used by astronauts or formed into fuel and oxidizer for propulsion, and is the subject of investigation for many Cislunar missions (TO20A, PRIME-1, TO19C). Operations of the South Pole are further enabled by the existence of the NRHO and the Gateway station contained within. The NRHO is a piece of the small set of stable orbits that exist in such close proximity of the Moon. Its passage close to the Moon, with a significant portion of its orbit above the South Pole, allows for it to act as a stepping stone between the Lunar surface and space, further enabling operations in the South Pole.} Fifteen missions are planned to attend the South Pole region, with many more likely to attend after the establishment of the Artemis base. The South Pole is a vital region of interest on the Lunar surface as it hosts the largest population of missions planned for the next decade, and contains concentrations of water to be utilized by Cislunar missions.

\subsubsection{Cislunar Space}
Analysis of missions attending the Lunar surface identifies important locations on the Moon, but does not encompass all key regions of space. A multitude of missions such as Artemis I, Artemis II, KPLO, Lunar IceCube, Lunar Trailblazer and others, are orbiters and will not land on the surface. Thus, identification of regions of interest in Cislunar space shall also be established to further characterize key portions of the Cislunar region. Two key traits are utilized to characterize regions of interest in Cislunar space: the population of missions in a region, and the region's viability to provide Cislunar SDA applications. First, by identifying the trajectories used by many Cislunar missions, orbits of interest are identified. This method is simple and insightful, but is challenging to complete for missions as agencies often {\color{black}do not share precise orbit information. General context on a mission trajectory is typically provided beforehand and is used to place these missions into their general orbit or orbit family.} Therefore, further analyzing regions of space utilized for observation, communication relay, surveillance, or other Cislunar space domain awareness services, provides a refined insight into Cislunar space regions of interests.

\begin{figure}[t!]
    \hfill{}
    \centering
    \includegraphics[width=13.5cm]{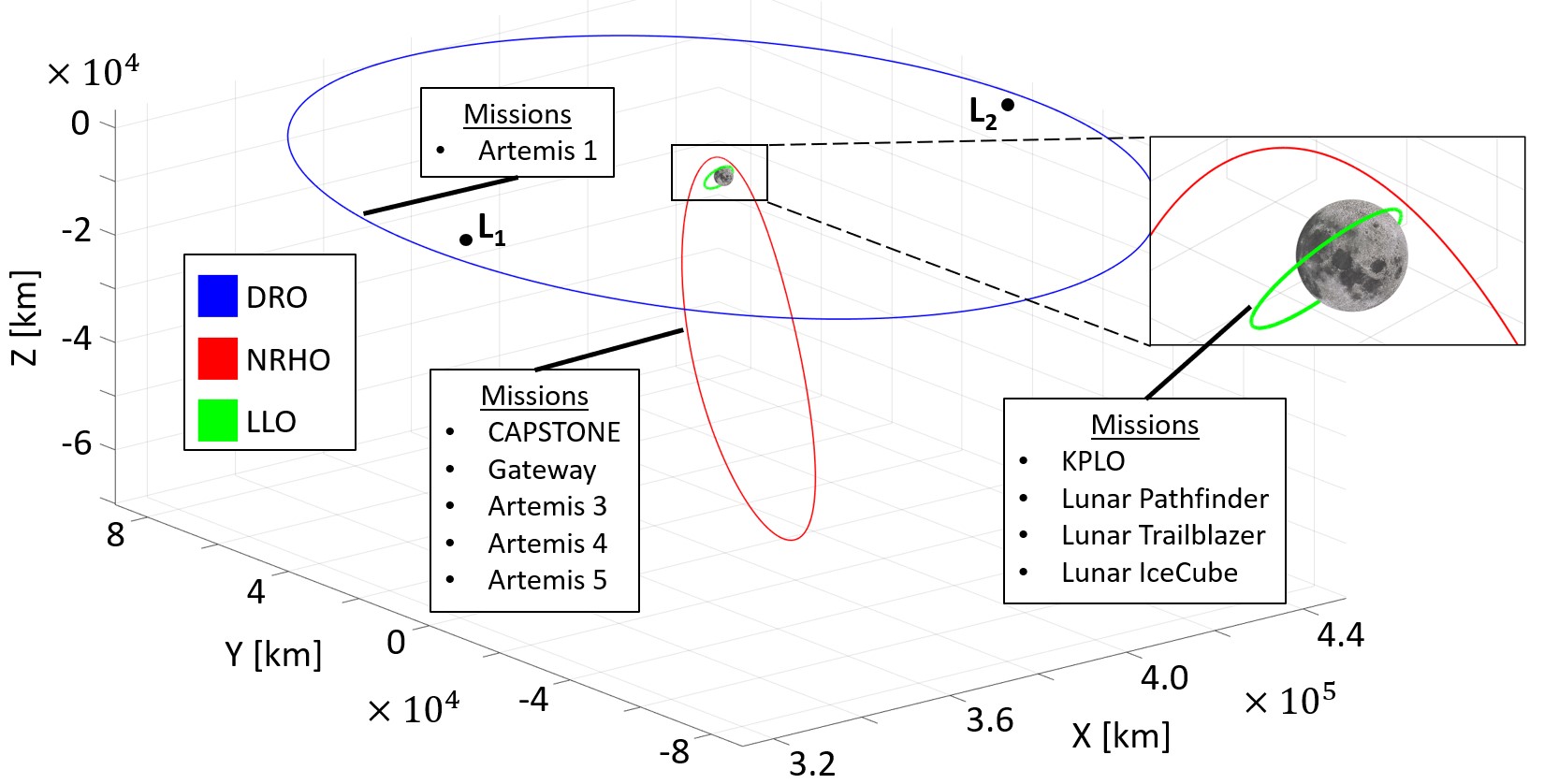}\hfill{}
    \caption{\label{fig:OrbitMap}Typical orbits of interest in the Cislunar region, presented in Earth-Moon barycenter rotating frame.}
\end{figure}

Out of the missions highlighted in Section \ref{sec:Missions}, a significant portion {\color{black}are contained close to the Moon or operate in LLO}. Referenced missions whose planned trajectories are identified are plotted in Figure \ref{fig:OrbitMap}. Note that a mission is categorized under a sample orbit similar to its true orbit, as the exact path is not always {\color{black}attainable and to help classify the motion.} Two regions of interest emerge from this plot: NRHO and LLO. The southern $L_2$ NRHO is the orbit where Gateway will reside for its lifetime. The trajectory offers two key features beneficial for Gateway and Artemis: a view of the South Pole and low station-keeping requirements to remain in orbit for long periods of time. Since Gateway is a pivotal piece of infrastructure in Cislunar space, missions docking onto Gateway must enter a similar NRHO. Furthermore, the NRHO is unique in that it passes extremely close to the North Pole, passing swiftly over, and then quickly ascends a large distance from the Moon, spending the majority of its orbit above the South Pole. However, the large distance from the Lunar surface is not advantageous for experiments that aim to precisely study the Lunar surface. Thus, the key of the NRHO is to provide communication with Earth and the South Pole as well as {\color{black}transportation from the orbit to the Lunar surface and vice versa.} 

\begin{figure}[t!]
    \hfill{}
    \centering
    \includegraphics[width=8cm]{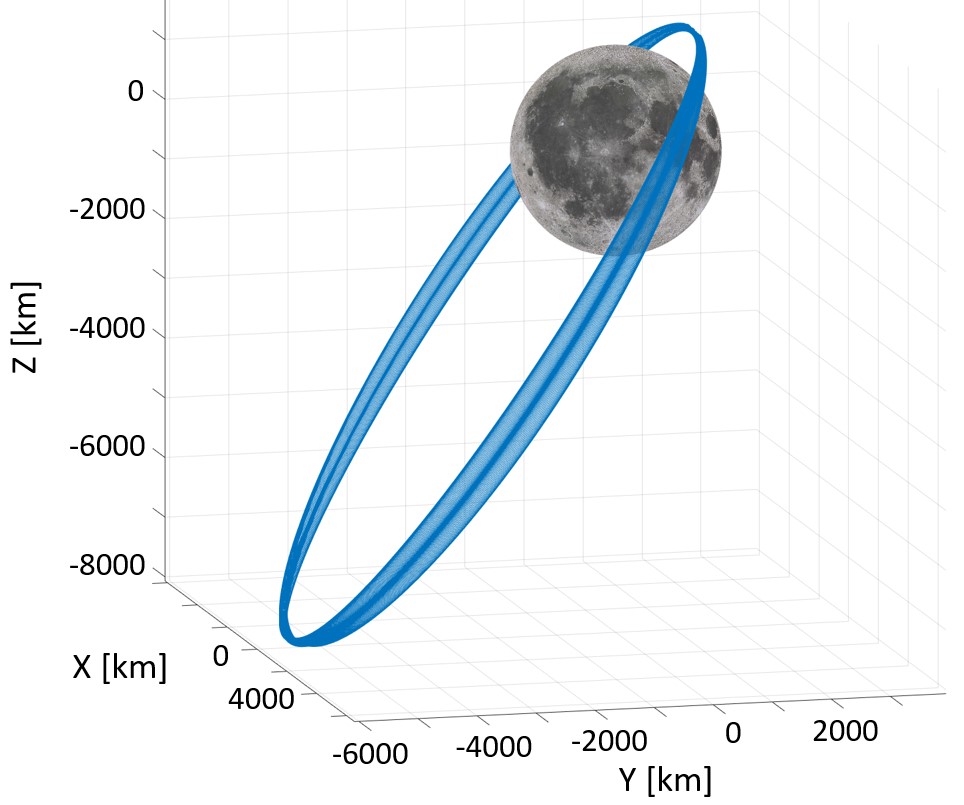}\hfill{}
    \caption{\label{fig:ELFO}{\color{black}Sample ELFO propagated for $4$ weeks, presented in a Moon centered, Earth-Moon rotating frame.}}
\end{figure}

{\color{black} Many Cislunar missions will operate about the Moon in LLOs.} Missions aim to utilize the lower altitudes for experiments to collect accurate data about the Lunar surface. {\color{black}Some of these missions also take advantage of a unique subset of LLOs, referred to as elliptical frozen Lunar orbits (ELFO). The ELFO oscillate slightly about the Moon but retain their shape and general orientation for long periods of time \cite{ELFOCon}. There are a couple sets of ELFO that surround the Moon and a single example is shown in Figure \ref{fig:ELFO}; note  the oscillatory motion but overall retained orientation and shape of the orbit.} Different studies into Lunar topography, water distribution~\cite{IceCube}, regolith composition, and communication relays~\cite{Pathfinder} to Earth will all be carried out in LLO. The low altitude also allows for the Moon to dominate the gravitational influence of a spacecraft, producing motion similar to conics. However, in the Earth-Moon system, the LLO trajectories are not precisely conic and will eventually lose its conic shape. {\color{black}Being only slightly perturbed by the presence of Earth's gravity,} LLO are of further interest because those that are contained in low altitudes and are highly inclined are sufficient to observe the Lunar South Pole.

Corresponding to the dynamics of the CR3BP, the libration points of the Earth-Moon system must also be considered key portions of Cislunar space. A large number of orbit families are dynamically related to a respective libration point, either orbiting the point or extending from a family that does. It is noted that the majority of missions traversing the Cislunar realm are specifically interested in the Moon. On that account, the $L_1$ and $L_2$ libration points and associated dynamics/families are most likely to be utilized by Cislunar missions. Figure \ref{fig:OrbitMap} depicts that the majority of missions are in trajectories falling between $L_1$ and $L_2$, with an $L_2$ NRHO being identified specifically. Taking a step back from the Moon, the Earth-Moon $L_4$ and $L_5$ points continue to have investigators researching their utility in science, communication, and surveillance~\cite{LunarOccultations}, and may see missions attending the area in the future.

\subsubsection{{\color{black} Challenges Facing Cislunar Missions}}
\begin{figure}[t!]
    \hfill{}
    \centering
    \includegraphics[width=15cm]{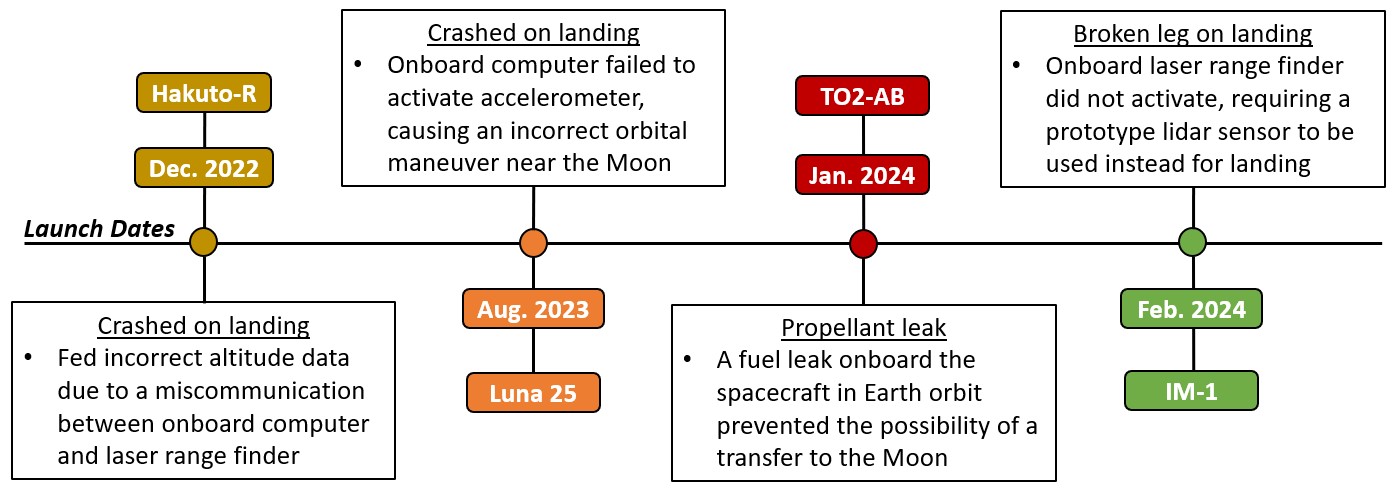}\hfill{}
    \caption{\label{fig:MissionFailure}{\color{black}Summary of failures for Cislunar missions that occurred in the early 2020s.}}
\end{figure}

{\color{black}A common hurdle faced by many Cislunar missions is the problem of landing. Across modern Cislunar missions, Channdrayan-2 in 2019, Hakuto-R 1 in 2022, and Luna 25 in 2023, successfully traversed to the Moon but faced failures during landing, causing destruction on impact. Additionally, the IM-1 completed a soft landing but fell over onto its side after a strut collapsed. {\color{black} A summary of recent Cislunar mission failures is shown in Figure \ref{fig:MissionFailure}}. From these failures, the struggle of landing onto the Moon is highlighted. Missions TO IM-1, TO 2-AB, TO PRIME-1, and TO 19C are carrying laser retroreflector arrays that provide laser ranging for the spacecraft to utilize during landing \cite{Cremons20}, {\color{black} in efforts to continue to manage successful Lunar landings.} Furthermore, to study landing plumes, the IM-1 and TO 19D mission carry stereo cameras to collect data during landing. Given these points, more missions are being proposed to enter into the region and test new technologies that aim to address significant challenges, such as landing or harvesting resources for sustainable operations. The commercial industry, along with large-scale programs such as Artemis and Chang'E, will cause traffic in Cislunar space to become consistent} {\color{black} and will continue to refine landing operations and technology}.

{\color{black} Inevitably, the future holds new challenges that will emerge with the continued expansion of the Cislunar region. However diverse these challenges are, the issues of sustainability will persist throughout the future of Cislunar space. Humanity aims to establish a long-term occupancy of humans in the region, a goal that necessitates sustainable operations. Strides are already being taken to promote sustainability, as seen by the establishment of Lunar bases, or ISRU technology being developed to harvest resources directly on the Moon. Even the selection of Gateway in the Southern NRHO, requiring small reserves of propellant to maintain \cite{GatewayNRHO}, further highlights the foundation of sustainable practices in Cislunar space to promote long-term operation. A reflection on a current challenge of sustainable space operations about the Earth, the removal and cataloguing of space debris, provides insight into potential challenges the Cislunar region may face in the future. The influx of Lunar missions in the near future will enhance the residence of space debris in the region. Furthermore, the presence of space debris in Cislunar space are potentially more hazardous to sensitive missions compared to operation and debris about the Earth. The natural dynamics of the Earth-Moon system posses the risk of debris to be trapped in the few stable regions that exist in Cislunar space, eliminating the ability to leverage these stable regions for future missions. Additionally, the majority of the instrumentation used in networks to measure, track, and catalogue debris in Earths proximity does not extend into Cislunar space, leaving debris untracked and dangerous to operations. This increase in space debris, in conjunction with the natural dynamics of the region, poses a significant risk to the sustainability of Cislunar missions in the future. The impacts of substantial quantities of space debris in Cislunar space have been analyzed \cite{Boone}, while current efforts toward debris mitigation are addressed in proposed policies for operations in the region \cite{CislunarDebris}. Cislunar expansion will face challenges pertaining to the sustainable operations of the region, such as sourcing resources or managing space debris, that must be overcome for the region to remain viable as a passage into deep space. These risks will be mitigated through the creation of policies for space sustainability to ensure long-term operational success.}

\section{Cislunar Space Domain Awareness} \label{sec:CislunarSDA}
Classically, Space Situational Awareness (SSA) refers to the comprehensive current and predictive knowledge of space objects and their operational environment. Cislunar Space Domain Awareness (CSDA) involves the essential knowledge needed for maintaining current space functions in the Cislunar space and determining reasons for any disruptions or losses ~\cite{holzinger2018challenges, Primer}. The purpose of CSDA is to provide actionable strategies prior to upcoming threats or hazards by tracking and predicting all functional and non-functional space objects to promote sustainable and safe use of Cislunar space. {\color{black}Some of the current CSDA efforts aim to investigate ways to establish surveillance, the detection of new objects, cataloging, initial orbit determination and the effects of space debris and fragmentation events in this region .}
\begin{table}[h!]
\caption{{\color{black}Summary of common sensors for SDA applications.}}
\label{SensorTable}
\begin{tabular}{lll}
\hline
\textbf{Sensor Type} & \textbf{Advantages}                             & \textbf{Drawbacks}                                      \\ \hline
Radar Sensors        & No dependency on illumination conditions. & High antenna temperature decreases SNR.            \\
\multicolumn{1}{c}{} & Measures range \& range rate  (if Doppler radar).   & High power consumption for distant objects.      \\
                     & Covers a volume of space within  variable beam width.                 &                                                    \\
                     &                                           &                                                    \\
Laser Sensors        & No dependency on illumination conditions. & Narrow beam. \\
                     & Measures range; eventually angles are reported.   &                           \\
\multicolumn{1}{c}{} & \multicolumn{1}{c}{}                      &                                                    \\
Optical Sensors      & Passive sensors.                          & No information on range or range rate.             \\
                     & Measures two angles.    & Requires proper illumination conditions.           \\
                     & Covers space volume commensurate with field of view.                 &                                                    \\
\multicolumn{1}{c}{} & Relatively cheap to operate.              &                                                    \\ \hline
\end{tabular}
\end{table}

Different sensors may be employed for detection and tracking; for example, standard sensors would include optical, radar, and laser-ranging observations (Table \ref{SensorTable}). Radar sensors are active sensors, i.e., they do not rely on the illumination of the object due to the Sun for detection. They typically measure range and eventually two pointing angles. Active sensors are range limited via the distance quadratic, further distances require beam-narrowing \cite{RadarRef}; traditionally radar has been used for  for low-Earth orbit region {\color{black}or short range applications}. The signal-to-noise ratio (SNR) for a fixed transmitted radar signal decreases as the antenna temperature increases. For the low Solar phase angles, the temperature of the antenna increases as it is pointing in the direction of the Sun, thereby hindering the detection of an object~\cite{ASC22_1}. Then, laser ranging sensors are likewise active sensors. They typically measure range and eventually two pointing angles; laser beam width are narrow, causing more stringent pointing requirements, and usually a retroreflector is required. Finally, optical sensors are passive sensors that rely on the object's illumination by the Sun to measure two angles. They are comparatively cheaper to operate than radar and laser ranging sensors, and have a much less demanding power demand. {\color{black}They are commonly used in ground observatories and see application on-board spacecraft, such as in star trackers that use electro-optical sensors. }

For optical sensors system, the observations are constrained by avoidance angles from the Sun, Moon, and Earth, the object's visual magnitude, and background light sources. For ground-based optical observations, the object has to be above the local horizon at sufficiently low sun light conditions, which usually result in a night time constraint. In order to model the optical brightness of a sun illuminated object at the site of the observer, the magnitude of a space object may be computed~\cite{fruehcislunar}: 
\begin{eqnarray}
\mathrm{mag} = \mathrm{mag_\mathrm{sun}} - 2.5 \log_{10}(\frac{I_\mathrm{sc}}{I_\mathrm{sun}})\label{mag}
\end{eqnarray}
where $\mathrm{mag_\mathrm{sun}}$ is the apparent reference magnitude of the Sun, $I_\mathrm{sc}$ is the irradiance reflected off the spacecraft and $I_{Sun}$ is the Sun's reference irradiance. In order to compute the irradiance that is reflected off the object, assumptions about the albedo area, the shape, and the attitude of the space object have to be made. {\color{black}The Lambertian reflection model used in Eq.\ref{mag} is often employed to model the reflection of light off of surfaces and objects. Lambertian reflectance indicates that the reflection of light off a surface has the same apparent magnitude regardless of viewing angle. For a spherical object with Lambertian reflection model,} the irradiance of the object is expressed via the following relation~\cite{CFamos19}: 
\begin{eqnarray}
I_\mathrm{sc} = \frac{I_\mathrm{Sun}}{d_\mathrm{sc-obs}^2}\frac{2}{3}\frac{C_d}{\pi}r^2(\sin\alpha+(\pi-\alpha)\cos\alpha)
\label{cd}
\end{eqnarray}
where $ d_\mathrm{sc-obs}$ is the distance between the object and the observer, $r$ is the object's radius, and $\alpha$ is the phase angle between the Sun and the observer at the object's location. {\color{black}The Lambertian diffuse reflection coefficient ($C_d$) is a constant that describes how a surface diffuses light, dependent upon surface color and material.} It is important to note that the solar constant $I_0$ needs to be scaled to the actual distance between the Sun and the object ($d_\mathrm{Sun-sc}$) for the reference irradiance $I_\mathrm{Sun}= I_0\frac{\mathrm{AU}^2}{d_\mathrm{Sun-sc}^2} $. This is due to the fact that the solar constant is defined at one Astronomical Unit (AU). {\color{black}The relationship between phase angle, diffuse reflection coefficient, and visual magnitude for a spherical object is shown in Figure \ref{fig:phasevsmag}. 
\begin{figure}[b!]
    \hfill{}
    \centering
    \includegraphics[width=0.55\textwidth]{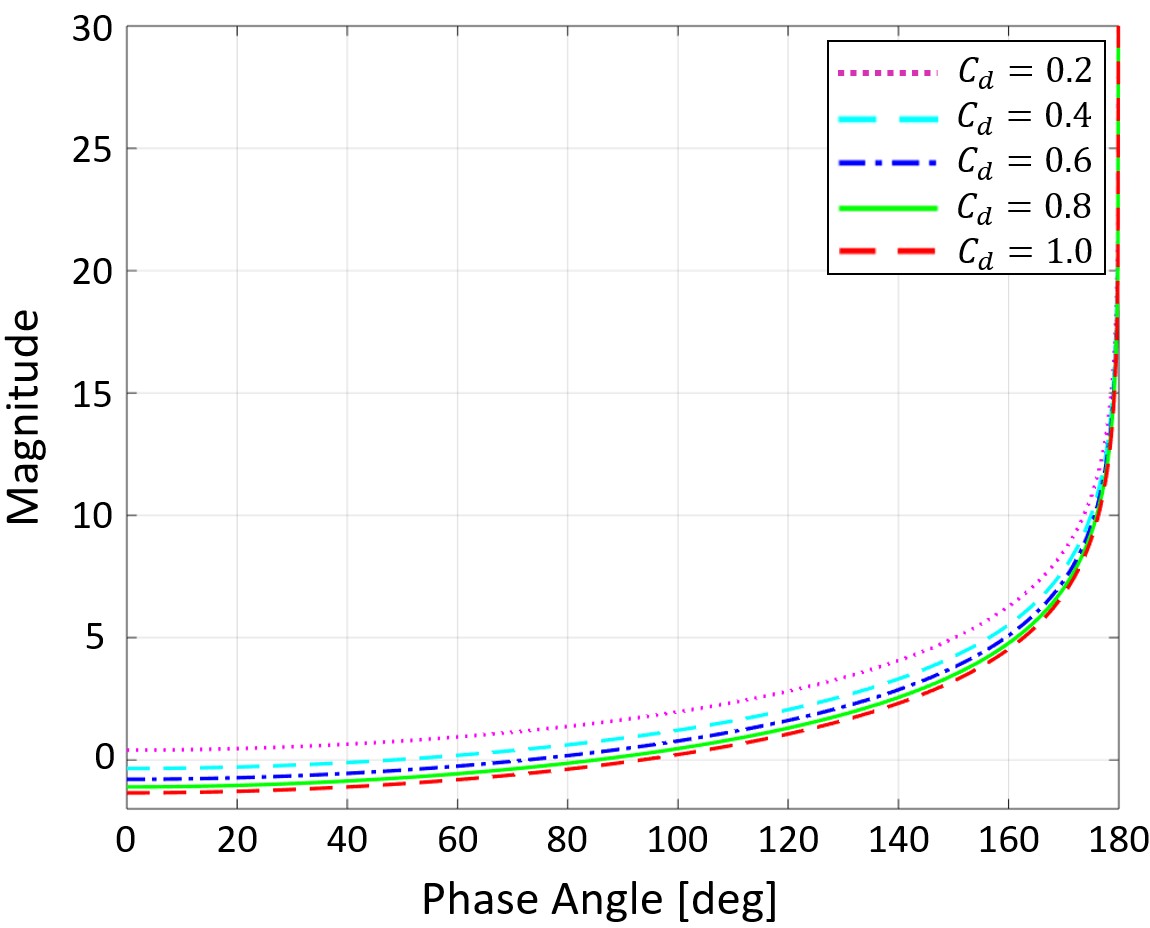}\hfill{}
    \caption{ {\color{black}Visual magnitude of a one meter sphere with varying reflectivity coefficients ($C_d$) across different phase angles.}}
    \label{fig:phasevsmag}
\end{figure}
The size of the object is one meter in radius and the reflectivity is varied from 0.1 to 1. As demonstrated, a higher reflectivity and low phase angles provides a low visual magnitude. The lower the visual magnitude, or the brighter an object, the more favorable the illumination conditions are for a possible observation.} Although the Lambertian reflection model for a spherical object overlooks glints, like those from a box-wing, which appear in specific geometries and primarily at minimal phase angles for sun-oriented satellites, it provides a baseline for average visibility conditions. Exact satellite models may always be inserted into any given scenario.

Previously, a couple of approaches have been explored to study CSDA. Holzinger et al.~\cite{holzinger2018challenges, Primer} gave the definition and purpose of CSDA along with the overview of dynamics, challenges in the observations, and spacecraft operation in Cislunar space. The initial publications in the field examined the viability of certain orbits for Cislunar applications. Fowler et al. \cite{fowler2020observability} performed visibility  analysis for a combination of target and observer trajectories based on their range, unavailability, and angular interval. Bolden et al.~\cite{bolden2020evaluation} explored a variety of orbits for SNR, orbital stability, and too-short-arc challenges in the Cislunar domain. Wilmer et al.~\cite{wilmer2021preliminary, wilmer2021cislunar, Dahlke} analyzed periodic orbits for their effectiveness when subjected to the SDA mission architecture of monitoring targets around $L_1$, $L_2$, and $L_3$ libration points. Frueh et al.~\cite{fruehcislunar,FruehSSA} investigated conditions for successful SDA by evaluating the visibilities of selected orbits and concluded that only ground-based sensors are not sufficient for surveying the orbits in the Cislunar regime. To expand the visibility analysis for planar and three-dimensional cases using space-based sensors, Bhadauria et al.~\cite{ASC22Surabhi, AMOSSurabhi} worked on the parameterization of the Cislunar space {\color{black}including the Sun's gravitational influence}, and produced feasible regions for surveillance using the visibility maps. Orbits in the CR3BP suitable for surveillance in Cislunar space have been presented with eclipse avoidance techniques that leverage the natural dynamics for trajectory design ~\cite{GuptaSSA,gupta2022long, Gupta2023eclipse}. A dual-use star tracker has also been tested in space to verify the performance and functional requirements of both star tracker and SDA sensor functionalities \cite{plotke2021dual}. The feasibility of GEO, HEO, and x-GEO\footnote{X-Geo is referred to all space beyond GEO if measured from Earth.}-based optical sensors has also been examined \cite{thornton2022dim, laone2022leveraging}. {\color{black} It has been concluded that increasing the Lunar exclusion angle beyond 3 degrees decreases the tracking up-time for Halo orbits \cite{thornton2022dim} and that larger orbits in the x-GEO region have higher SNR due to larger observation times \cite{laone2022leveraging}. A combination of ground-based and space-based sensors for SDA applications has been surveyed to show that space-based sensors mitigate many of the ground-based sensor limitations like solar exclusion, weather outages and restricted geographical locations \cite{bloom2022space}.}  DeCoster et al.~\cite{decoster2023orbit} presented a paradigm shift that enables the manufacturing of large apertures in situ or on the Lunar surface for a combination of large-scale RF and optical sensors. The design of a Cislunar satellite constellation has been provided by Patel et al.~\cite{Patel2023constellation} based on the phasing and the number of observing satellites. The ongoing research in CSDA visibility is widely exploring the feasibility of ground-based and space-based observers for effectively surveying the Cislunar space.

For fragmentation analysis in Cislunar space, previous researchers investigated the explosion of operational satellites in the collinear ($L_1$, $L_2$, $L_3$) Sun-Earth Libration points. Findings from the paper indicated that half of the fragments drifted toward the Earth, and the other half drifted away from it~\cite{landgraf2001space}. Boone et al., Bettinger et al., and Wilmer et al. studied the movement of space debris from major spacecraft accidents near Earth-Moon Lagrange points $L_1$, $L_2$, $L_4$, and $L_5$, as well as in Cislunar periodic orbits, polar Lunar orbits, LLO, and Lunar interplanetary dust grains close to the stable Earth-Moon $L_4$ and $L_5$ Lagrange points ~\cite{boone2021debris, wilmer2021artificial, boone2021spacecraft, boone2021artificial, bettingersurvivability, boone2022long, bettinger2021spacecraft, wilmer2022debris}. Guardabasso et al.~\cite{guardabasso2023analysis} studies fragmentation for a southern NRHO about the second Lagrange point of Earth–Moon system and a DRO using the NASA Standard Breakup Model to lay the foundation for the Cislunar Space Debris Mitigation framework. Black et al.~\cite{black2023characterizing, black2023investigation} used a modified version of the NASA Standard Breakup Model to simulate spacecraft fragmentations in $L_1$ and $L_2$ Lyapunov orbits, and then investigated the fragment behavior as a function of energy level and orbit location in the vicinity of the $L_2$ Lagrange point.

\begin{table}[b!]
\caption{{\color{black}List of orbits/regions analyzed for SDA fragmentation events.}}
\label{fragmentationEvents}
\begin{tabular}{lll}
\hline
\textbf{Orbits/Regions} &                                    \\ \hline
Sun-Earth Lagrange Points
($L_1$, $L_2$, $L_3$)      \\
Earth-Moon Lagrange Points 
($L_1$, $L_2$, $L_4$, and $L_5$)     \\    
Cislunar Periodic Orbits\\
Polar Lunar Orbits \\
Low-Lunar Orbit (LLO) \\
$L_2$ Southern NRHO \\
DRO \\
$L_1$ and $L_2$ Lyapunov Orbits \\
 \hline
\end{tabular}
\end{table}

The established SSA methods in detection and tracking, initial orbit determination, and navigation fail due to the shift from the two-body dynamics to multi-body dynamics in Cislunar space. Therefore, new methods are being developed to catalog space objects successfully. Authors  have used invariant manifold structures to define highways where a lost spacecraft must reside to transit to different regimes ~\cite{ASC22_4, hall2022reachability}. Additionally, a Taylor series approximation to reachability set based on maximum thrust and a window for the time of maneuver has been defined. Kabir et al.~\cite{Kabir2023unknown} worked on detecting an unknown, non-cooperative, and dynamic space object by a chaser satellite based on a Bayesian optimization framework. Priedhorsky et al.~\cite{Priedhorsky2023detect} modeled the detectability of Cislunar space objects without a priori knowledge from a space-based telescope. The initial orbit determination in the Cislunar space has been explored by~\cite{thompson2021cislunar, ii2021cislunar,ii2022cislunar, wishnek2021robust, ASC22_3, ASC22_5}. For navigation, Mercurio et al.~\cite{Mercurio2023relnav} worked on relative navigation in Cislunar space based on a modal representation of relative motion and derived a filtering framework for relative navigation in CR3BP. Givens et al.~\cite{Givens2023nav} proposed using a Gaussian Mixture Model (GMM) filter with visual direction-of-motion measurement and studied its efficacy with respect to EKF. Hippelheuser et al.~\cite{Hippelheuser2023EKFUKF} compared the extended Kalman filter with the unscented Kalman filter to utilize observation from space-based observers and concluded that for the small measurement step size, EKF would give accurate results and low computation cost, and for large measurement step size, UKF would provide great accuracy for no extra computational cost. Khatri et al.~\cite{Khatri2023uncertainty} leveraged the GMMs and State Transition Tensors in the CR3BP for accurate uncertainty propagation and further used the GMMs to compute the probability of collision of two objects. Other related areas which have been explored in the Cislunar domain include estimation and uncertainty~\cite{greaves2021relative, wolf2022multi}, sensor tasking~\cite{fedeler2022sensor, harris2021expanding, ASC22_1} and Cislunar constellation~\cite{Peng2023nav, Hartigan2023nav}. 

{\color{black}The SDA limitations and current observational and orbit determination gap in Cislunar space hinders the safe operation and expansion of the region for the many missions attending over the upcoming decades. The majority of the work reviewed in this section aims to sew the gap through the application of sensor networks throughout the region as well as apply new navigation and monitoring techniques. Optical sensor applications are at the forefront of sensor applications due to their passive nature and long range capabilities. These sensors are continuously challenged by the requirement of proper target illumination while avoiding the illumination from other celestial bodies, resulting in the need for strategic placement \cite{bolden2020evaluation, wilmer2021preliminary, wilmer2021cislunar, Dahlke, ASC22Surabhi, AMOSSurabhi, ObsSc}. Radar sensors are able to operate without the need to satisfy illumination conditions, but lack real application due to the power requirements needed to span the vast region of Cislunar space. However, these sensors possess the potential for near Moon applications over short distances. Leaving key pieces of Cislunar space without proper space domain awareness capabilities, enabled by these sensors, imposes potential risks for operations in the region. Potential collisions with untracked space debris is a hazard that will only continue to grow as missions consistently traverse the area and as fragmentations occur. The continued work to understand  the natural motion and tracking of Cislunar objects is necessary to mitigate the risk of collisions \cite{guardabasso2021lunar, black2023characterizing, black2023investigation, ASC22_4, hall2022reachability}. Overall, for the continued success and health of Cislunar missions, Cislunar SDA weaknesses such as, methods of observing objects, tracking debris, and precise navigation for missions throughout the region, must continue to be improved.}

\begin{figure}[t!]
    \hfill{}
    \centering
    \includegraphics[width=0.7\textwidth]{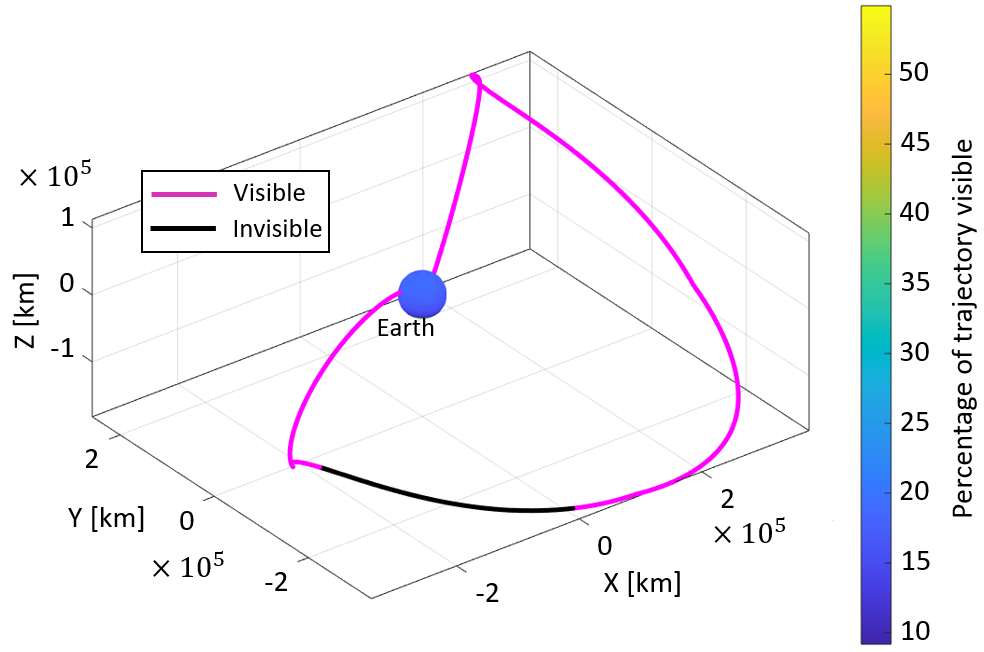}\hfill{}
    \caption{{\color{black}Visibility of Artemis I's trajectory viewed from a network of optical sensors on Earth, 11/16/2022- 12/11/2022, presented in the J2000 inertial frame (Earth not to scale).}}
    \label{fig:L_visglobal}
\end{figure}

{\color{black}Artemis I was launched on November 16, 2022 with the purpose of testing the Orion capsule's operational capabilities and ability to safely carry a crew. Artemis I's trajectory spans about 25 days and its orbit data is pulled from NASA JPL's Horizons system \cite{Horizons}, a system that catalogs and stores ephemeris data of planets, moons, spacecraft, and other objects. The Artemis I trajectory is used here as an example to demonstrate the applications of optical observation techniques and the limitations of these current techniques in Cislunar space (Figure \ref{fig:L_visglobal}).}
{\color{black}A network of optical sensors in a 20-by-20 grid is configured on the surface of the Earth. The spacing of the 400 sensors is 9.4 degrees in latitude and 18.9 degrees in longitude around the Earth. Artemis I is deemed observable by a sensor if the Orion's visual magnitude is at most 20 and if it is brighter than the magnitude of the moonlight. Recall, the smaller the visual magnitude of an object, the brighter an object is. Furthermore, the capsule must be above a sensor's local horizon on the Earth's surface during the night to be considered visible. In other words, Orion must be in a sensor's night sky to be potentially visible. In order to model the Orion capsule and the light that it reflects to the optical sensors, its shape, size, and reflectivity are selected. In reality, the Orion capsule is about 7.3 meter in height and 5.2 meter in diameter and mostly white and silver. In combination of simplicity and accurate modeling, the Orion capsule is assumed to be a sphere of diameter 5.0 $m$ and reflectivity of 0.5. The simulation of of Artemis I is then completed in Figure \ref{fig:L_visglobal} using the described visibility conditions, sensor network, and Orion model. If all visibility conditions are met for at least one sensor on the Earth, Orion is considered visible and the corresponding portion of the trajectory at that time is colored in magenta in Figure \ref{fig:L_visglobal}. Otherwise if it is not visible, the trajectory is colored black. In the figure, the Earth is presented as a sphere broken into grids, where at the center of each grid lays a sensor. A sensor's grid is colored using the colorbar based upon the duration that the sensor is able to view Artemis I through the simulation, parameterized as a percentage of Artemis I's entire trajectory.}

{\color{black}The resulting simulation of a global ground-based optical network observing the Artemis I mission demonstrates the difficulties of providing SDA in Cislunar space even if the sensors are spread globally. A notable portion of the trajectory (about 9 days) is not observable from this network. Furthermore, this time span is not intermittent but rather one continuous gap in coverage. Looking to the Earth, the majority of the sensors fell below a $20\%$ view of the total Artemis I trajectory. {\color{black} Thus, in a more realistic scenario with a significantly reduced network of sensors, it poses a significant challenge to observe objects in Cislunar space.} Additionally, most debris are smaller in size than a five meter diameter object like the Orion capsule, so observing these objects becomes a daunting task. Finally, Artemis I is a high profile mission and sways the attention of cutting edge observational systems, such as the Deep Space Network, to provide any possible tracking and navigation services all the way out in Cislunar space. Unfortunately, for the many objects, debris, or smaller scale missions that do not receive the same priority or attention as Artemis I, a global network of observational sensors will not be able to provide sufficient SDA capabilities in Cislunar space, as demonstrated by this analysis. Through the development of Cislunar SDA techniques, in particular focused in the key regions of interests described in Section \ref{sec:ROI}, the numerous other missions traversing Cislunar space may operate in a safe and monitored environment.}

{\color{black} The state of the art in Cislunar space domain awareness continues to evolve rapidly, driven by advancements in technology, increasing interest in Lunar exploration and utilization, and the growing importance of space situational awareness. There have been significant advancements in sensor technology, including improved space-based telescopes, radar systems, and optical sensors, enhancing the capability to detect, track, and characterize objects in Cislunar space with higher precision and sensitivity. For the persistent success of Lunar missions, SDA techniques must continue to branch out into Cislunar space, as seen by the efforts described in this section.}

\section{Conclusion} \label{sec:Conclusion}
The Cislunar region is the focus of human expansion into space over the next decade. Numerous missions are attending the region, bringing an assortment of payloads and experiments that will expand humanities' knowledge of the Moon and space. Many of these missions aim to build the infrastructure needed to further support the long-term presence of humans on the Moon. Infrastructure such as Lunar space and ground stations will act as a foothold, increasing the longevity of Lunar missions and providing a staging point for launching towards Mars and deep space. At the forefront of Cislunar expansion, the United States pushes into the region through the Artemis program and Gateway, while enabling the commercial space industry's roll through the CLPS program. Other countries and organization also pursue a dominate Cislunar presence, further adding to the increasing volume of traffic. These missions do not distribute evenly across Cislunar space and on the Lunar surface, and instead focus on key regions of interest. On the Moon, the Lunar South Pole and Earth-Sided craters see a concentration of Lunar missions. While in Cislunar space, NRHO, DRO, and LLO house the majority of Cislunar on-orbit missions. As the traffic continuously concentrates in these regions, the need to monitor and track missions as well as all objects near these key regions grows. Achieving space domain awareness in Cislunar space is challenging. {\color{black}The application of Earth-based sensors used to improve SDA about the Earth do not suffice for Cislunar space, as shown by the Artemis I analysis. However, improved Cislunar SDA is attainable using sensors like optical, radar, and laser-ranging, complemented by innovative techniques being explored. Overall, a comprehensive review of the Cislunar region is presented containing: (a) a synthesis of modern Cislunar missions, their motivations, and where they are settling, and (b) an in-depth summary of Cislunar SDA techniques, available sensors, and the limitations of current SDA applications to Cislunar space. In synthesizing these findings, it becomes evident that as humanity ventures further into the Cislunar domain, a harmonious integration of strategic mission planning and vigilant space domain awareness} will be pivotal for the success of future space endeavors.

\section{CRediT Author Statement}
\noindent \textbf{Brian Baker-McEvilly:} Conceptualization, software, formal analysis, investigation, writing-original draft \\
\textbf{Surabhi Bhadauria:} Investigation, writing-original draft, formal analysis  \\
\textbf{David Canales:}  Conceptualization, supervision, resources, writing-review \& editing \\
\textbf{Carolin Frueh:}  Conceptualization, supervision, writing-review \& editing

\bibliographystyle{citationstyle} 
\bibliography{refs}

\clearpage
\appendix

\end{document}